\documentclass[a4paper,11pt]{article}
\pdfoutput=1
\usepackage{jheppub}
\usepackage{amsmath}
\usepackage{amssymb}
\usepackage{color}
\usepackage{graphicx}
\usepackage{hyperref}
\usepackage{xspace,slashed}
\usepackage[utf8]{inputenc}
\usepackage{multirow}
\usepackage{algorithm}
\usepackage{algorithmic}
\usepackage{stackengine}
\usepackage{enumitem}
\usepackage{etoolbox}
\usepackage{mathrsfs}
\usepackage{multirow}
\usepackage{array}
\usepackage{physics}
\usepackage{framed}
\usepackage{bbold}

\newtoggle{draft}

\togglefalse{draft}

\iftoggle{draft} {
  \usepackage{changes}
}{%
  \usepackage[final]{changes}
}

\definechangesauthor[name=pn, color=blue]{pn}
\definechangesauthor[name=km, color=magenta]{km}

\usepackage{booktabs}

\makeatletter
\g@addto@macro\bfseries{\boldmath}
\makeatother

\newcommand{\cL}{\mathcal{L}}

\newcommand{\cO}{\mathcal{O}}

\newcommand{\ktot}{k_{\text{tot}}}

\newcommand{\rmd}{\mathrm{d}}

\renewcommand{\sec}[1]{Section~\ref{sec:#1}}

\newcommand{\app}[1]{Appendix~\ref{app:#1}}

\newcommand{\eq}[1]{Eq.~\eqref{eq:#1}}

\newcommand{\eps}{\epsilon}
\newcommand{\ep}{\epsilon}

\newcommand{\df}{\mathrm{d}}

\newcommand{\LT}{\mathscr{L}}

\newcommand{\tvar}{\ell}
\newcommand{\Tau}{{\cal T}_N}
\newcommand{\TauN}[1]{{\cal T}_{#1}}
\newcommand{\TauNinc}[1]{{\cal T}^{\rm inc.}_{#1}}
\newcommand{\TauZeroBar}{{\cal T}^{\rm inc.}_0}
\newcommand{\TauZero}{{\cal T}_0}
\newcommand{\TauBar}{{\cal T}^{\rm inc.}_N}
\newcommand{\TauHat}{\hat{\cal T}_N}
\newcommand{\colT}[1]{{\bf T}_{#1}}

\iftoggle{draft} {
  \newcommand{\tobedeleted}[1]{\textcolor{azure}{#1}}
  
}{%
  \newcommand{\tobedeleted}[1]{}
  
}

\newcommand{\be}{\begin{equation}}
\newcommand{\ee}{\end{equation}}

\usepackage{soul}
\allowdisplaybreaks

\newcommand{\NNNLO}{N$^3$LO}

\newcommand{\beq}{\begin{equation}}
\newcommand{\eeq}{\end{equation}}
\newcommand{\nn}{\nonumber}

\definecolor{azure}{rgb}{0.0, 0.9, 1.0}

\newcommand{\red}[1]{{\textcolor{red}{#1}}}

\newcommand{\TTPaff}{Institute for Theoretical Particle Physics,
  KIT, 76128 Karlsruhe, Germany}
\newcommand{\CERNaff}{CERN, Theoretical Physics Department, CH-1211 Geneva 23,
Switzerland}
\newcommand{\EDBaff}{Higgs Centre for Theoretical Physics,
School of Physics and Astronomy,
The University of Edinburgh, Edinburgh EH9 3FD, Scotland, UK}

\preprint{
  \begin {flushright}
    TTP26-012, P3H-26-027, CERN-TH-2026-055
  \end{flushright}
}

\title{$N$-Jettiness Soft Functions Made Simple}
\author[a]{Luca~Buonocore,}
\author[a]{Maximilian~Delto,}
\author[b]{Kirill~Melnikov,}
\author[a]{Pier~Francesco~Monni,}
\author[c]{Andrey~Pikelner,}
\author[a]{and Gherardo~Vita}

\affiliation[a]{\CERNaff}
\affiliation[b]{\TTPaff}
\affiliation[c]{\EDBaff}

\abstract{ We present a new method to compute the soft function for the
  $N$-Jettiness variable for arbitrary $N$ at high perturbative orders in QCD.
  It is based on the observation that the most singular part of the soft
  function, the dipole contribution, can be represented by a sum of an
  analytically calculable inclusive soft function and a remainder. The latter is
  absent at NLO, is immediately finite at NNLO, and we expect that can be made
  finite with the help of simple NLO-like infrared subtractions at \NNNLO. As a
  byproduct of this approach, we derive a very simple formula for the tripole
  contribution to the $N$-Jettiness NNLO soft function, which results in a fast
  numerical evaluation. We apply this method to compute the $N$-Jettiness soft
  function at NNLO, and report numerical results for up to five jets for the
  hadron-collider soft function. We finally outline the prospects for
  applications at \NNNLO. }

\begin{document}
\maketitle

\section{Introduction}\label{sec:intro}

The remarkable performance of the Large Hadron Collider (LHC) has ushered in an era where measurements of Standard Model observables are approaching percent-level precision~\cite{CMS:2019raw,ATLAS:2019zci,CMS:2021xjt,ATLAS:2022hro,ATLAS:2024nrd}. This trend will intensify at the High Luminosity LHC (HL-LHC), where the enormous increase in recorded number of events
and related decrease in statistical uncertainties of measurements
demands theoretical predictions of comparable accuracy to fully exploit the collider's physics potential~\cite{ATLAS:2019mfr,ATLAS:2022hsp,CMS:2025hfp}.

Significant progress has been made in extending QCD predictions to N$^3$LO for several processes where a color-singlet final state is produced~\cite{Anastasiou:2016cez,Mistlberger:2018etf,Duhr:2019kwi,Chen:2022cgv,Baglio:2022wzu,Campbell:2023lcy,Neumann:2022lft,Chen:2022lwc}. Although these calculations represent only initial steps toward general N$^3$LO accuracy at the LHC, they demonstrate that N$^3$LO QCD corrections can induce percent-level or even larger shifts in theoretical predictions for cross sections of key processes such as Higgs production via gluon fusion and Drell-Yan production. As HL-LHC measurements enter the percent regime across a broad range of processes, including these corrections becomes essential. Furthermore, N$^3$LO
QCD effects introduce new color correlations in the singularity structure of jet processes, potentially giving rise to genuinely new phenomena, such as factorization-breaking effects. Developing systematic methods capable of delivering fully differential N$^3$LO predictions is, therefore, an important goal for theoretical collider physics~\cite{Caola:2022ayt}.

A powerful strategy for obtaining fully differential fixed-order QCD predictions at hadron colliders relies on slicing or non-local subtraction methods~\cite{Catani:2007vq,Gaunt:2015pea,Boughezal:2015dva,Buonocore:2022mle,Abreu:2022zgo,Buonocore:2023rdw,Fu:2024fgj,Buonocore:2025ucn,Dong:2026uvw}. These approaches employ a physical resolution variable to isolate infrared (IR) sensitive regions of phase space and use associated factorization theorems to account for contributions of these regions analytically at high perturbative orders. Prominent examples include $q_T$~\cite{Catani:2007vq} and $N$-Jettiness~\cite{Boughezal:2015dva,Gaunt:2015pea} slicing, which have already yielded important results.
Indeed, $q_T$ slicing has enabled fully differential predictions for the Drell-Yan process~\cite{Chen:2022cgv,Campbell:2023lcy,Neumann:2022lft,Chen:2022lwc}, and all ingredients for applying $0$-Jettiness  slicing to color-singlet processes, such as Higgs and Drell-Yan production, are now available following the computation of three-loop beam~\cite{Ebert:2020unb,Baranowski:2022vcn} and soft functions~\cite{Baranowski:2022khd,Chen:2022yre,Baranowski:2024ene,Baranowski:2024vxg,Baranowski:2024ysi}, as well as logarithmically-enhanced power corrections~\cite{Vita:2024ypr}.

Extending these techniques to processes with jets is considerably more challenging. For processes where a color-singlet system is produced in association with a jet, NNLO QCD predictions have been obtained using $N$-Jettiness slicing~\cite{Boughezal:2015aha,Campbell:2023lcy,Alioli:2025hpa}, making it currently the only slicing method applicable to processes with jets. Moreover, efforts to extend $q_T$-like subtraction to final states with jets~\cite{Buonocore:2022mle,Buonocore:2023rdw,Fu:2024fgj,Buonocore:2025ysd} remain at an exploratory stage and are not yet practical to tackle N$^3$LO predictions. Given the importance of jet production at the LHC, developing robust, generalizable slicing methods for these processes is a critical task.

A major bottleneck in applying $N$-Jettiness slicing to generic processes at N$^3$LO is the calculation of soft functions involving an arbitrary number of light-like directions. Even for color-singlet production, the determination of the three-loop soft function with two Wilson lines required years of sustained effort and development of sophisticated multi-loop techniques~\cite{Baranowski:2021gxe,Baranowski:2020xlp,Baranowski:2022khd,Chen:2022yre,Baranowski:2024ene,Baranowski:2024vxg,Baranowski:2024ysi}. New ideas enabled recent computations of the NNLO $N$-Jettiness  soft function for arbitrary $N$~\cite{Bell:2023yso,Agarwal:2024gws}, but these computations also made it  clear that their extension  to  the next perturbative order is, at best, challenging.

The goal of this paper is to introduce a method that dramatically simplifies the calculation of $N$-Jettiness soft functions at higher orders.  The central point of this method is to isolate
 the dipole contribution for an arbitrary number of hard emitters and express it as the sum of an analytically computable piece and a finite remainder. Crucially, the remainder can be evaluated numerically with drastically reduced complexity. At NNLO, the remainder requires no infrared (IR) subtraction, and at N$^3$LO its structure resembles an NLO cross-section computation, requiring only NLO-type subtractions. We apply this method to the $N$-Jettiness soft function and present results at NNLO for general $N$ (and numerically up to $N=5$), demonstrating the power and flexibility of our approach.

The remainder of this paper is organized as follows. In Section~\ref{sec:Setup} we review the structure of $N$-Jettiness soft functions. We show how their divergences are captured by a simpler quantity, the \emph{inclusive} soft function, and explain how to use this observation to dramatically simplify the computation the $N$-Jettiness soft function at higher orders.
In Section~\ref{sec:0Jettiness}, we use this method to calculate the $0$-Jettiness soft function at NNLO. In Section~\ref{sec:NJettiness}, we extend this analysis to general $N$ and provide numerical benchmarks for selected phase-space points. We conclude in Section~\ref{sec:Conclusions}. Additional technical details are provided in appendices.

\section{Presentation of the method}
\label{sec:Setup}
%
The focus of this article is the soft function for the $N$-Jettiness variable defined in Ref.~\cite{Stewart:2010tn}. At the lowest order, we consider the scattering of $N+2$ resolved QCD partons that we denote by $h_i$. Throughout the paper, we use the notation
\be
\begin{split}
& 0 \to h_1 + h_2 + h_3 + h_4 + \cdots h_{N+2},
\\
& h_1  \to h_2+h_3  + h_4 + \cdots h_{N+2},
\\
&  h_1 + h_2 \to h_3 + h_4 + \cdots h_{N+2},
\label{eq:2.1}
\end{split}
\ee
so that the $N=0$ case always refers to a process with two color-charged particles at the Born level.\footnote{A different convention is often found in the literature where, for instance, the thrust-like variable in $e^+e^-$ is referred to as $2$-Jettiness.} The first process refers to $e^+e^-$ annihilation, the second to deep-inelastic scattering and the third to hadron collisions. Each parton in Eq.~(\ref{eq:2.1}) carries a four-momentum $p_i$ and a color charge $\colT{i}$.\footnote{Throughout this paper, we use the color-space formalism of Ref.~\cite{Catani:1996vz}.} All partons are assumed to be massless.
To define the soft function, we denote by $n_i^\mu\equiv p_i^\mu/p_i^0$ the light-like vector aligned with the direction of flight of the parton $h_i$.

The following discussion takes as a starting point the factorization theorem of Ref.~\cite{Stewart:2010tn}. More recently, an alternative factorization theorem for the $N$-Jettiness variable was presented in Ref.~\cite{Becher:2026kbr} to capture, in addition to the modes described by that of Ref.~\cite{Stewart:2010tn}, also the effect of Glauber modes. Concretely, these modes result in the appearance of coherence-violating-logarithmic (CVL) corrections at subleading color, which have been known to affect the resummation of scattering observables in the hadron-collider case~\cite{Forshaw:2006fk,Forshaw:2008cq,Forshaw:2009fz,Banfi:2010xy,DuranDelgado:2011tp,Gaunt:2014ska,Zeng:2015iba,Forshaw:2021fxs,Becher:2021zkk,Becher:2023mtx,Boer:2023jsy,Boer:2023ljq,Boer:2024hzh,Becher:2024nqc,Banfi:2025mra,Dasgupta:2025cgl,Becher:2026kbr}. Before proceeding, we note that the conclusions of Ref.~\cite{Becher:2026kbr} do not affect the results and the methods presented in this article. Specifically, the class of CVL corrections to the $N$-Jettiness factorization discussed in Ref.~\cite{Becher:2026kbr} start at N$^4$LO, and therefore, up to N$^3$LO, the predictions of the factorization theorems of Refs~\cite{Stewart:2010tn} and~\cite{Becher:2026kbr} are identical provided the correct direction of the Wilson lines is adopted.\footnote{We note that the CVL tower starts at N$^3$LO, but the first term is captured by the soft function defined in Eq.~\eqref{eq:soft_def}.}

\subsection{Definition and properties of the soft function}
\label{sec:def-props-SF}
In Soft-Collinear Effective Theory (SCET)~\cite{Bauer:2000ew, Bauer:2000yr, Bauer:2001ct, Bauer:2001yt}, the bare soft function is defined by the following operator matrix element
\begin{equation}\label{eq:soft_def}
    {\boldsymbol  {\cal S}}_{\Tau} (\tvar) = \langle
  0 | T\{{\boldsymbol Y}_{n_1} {\boldsymbol Y}_{n_2}\dots {\boldsymbol Y}_{n_{N+2}}\} \,\delta(\tvar-\TauHat)\,
  \bar{T}\{{\boldsymbol Y}^\dagger_{n_{N+2}} \dots {\boldsymbol Y}^\dagger_{n_{2}}{\boldsymbol Y}^\dagger_{n_{1}}\}| 0 \rangle\,,
\end{equation}
where the soft Wilson line $Y_n$ is expressed in terms of the soft gauge
field ${\bf A}^{(s)}_i$ as~\cite{Bauer:2001yt}
\begin{equation}
{\boldsymbol Y}_{n_i}(x) = {\cal P}\exp\left\{i g_s\int dt\, n_i\cdot
    {\bf A}_i^{(s)}(x+n_i t)\right\}\,,
\end{equation}
and analogously for $Y_{\bar n}$. The integral over $dt$ runs from $(-\infty,0]$ if $h_i$ is in the initial state and from $[0,+\infty)$ if $h_i$ is in the final state.
Here ${\bf A}_i^{(s)} \equiv {\bf T}_i\cdot A^{(s)}$, where ${\bf T}_i$ is the color charge of parton $h_i$.
The operator $\TauHat$ acts on a given state
of $X$ soft particles $|X_s\rangle$ with momenta $k^\mu_1,\dots,k^\mu_X$ by measuring the value of the $N$-Jettiness $\Tau$ of the corresponding final-state, namely
\begin{equation}
   \delta(\tvar-\TauHat) |X_s\rangle = \delta(\tvar-\Tau(k_1,k_2,\dots,k_X)) |X_s\rangle\,,
\end{equation}
where
\begin{equation}\label{eq:observable}
    \Tau(k_1,k_2,\dots,k_X) = \sum_{i=1}^X \min\{k_i\cdot n_1,k_i\cdot n_2,\dots,k_i\cdot n_{N+2}\}\,.
\end{equation}

The soft function in Eq.~\eqref{eq:soft_def} is conveniently studied in Laplace space; that is
\begin{equation}
 \widetilde {\boldsymbol  {\cal S}}_{\Tau} (u) = \int_0^\infty \rmd \tvar \, e^{-u\,\tvar}  {\boldsymbol  {\cal S}}_{\Tau}(\tvar) \,.
\end{equation}
In Laplace space, the soft function admits a local, multiplicative renormalization of the UV singularities in terms of the renormalization color matrix
 ${\boldsymbol Z}$ as
\begin{equation}\label{eq:renorm}
\widetilde {\boldsymbol  {\cal S}}_{\Tau} (u) =    {\boldsymbol Z} \widetilde {\boldsymbol  {\cal S}}^{(R)}_{\Tau} (u) {\boldsymbol Z} ^\dagger\,.
\end{equation}
From now on, unless necessary to clarify the notation, we will suppress the $u$ dependence in the Laplace transform of the soft function. 
The color matrices ${\boldsymbol Z}$ and $\widetilde {\boldsymbol  {\cal S}}_{\Tau}$ admit a perturbative expansion in powers of the bare strong coupling $\alpha_s$
\begin{equation}\label{eq:expansions}
   {\boldsymbol Z} = {\mathbb 1} + \sum_{n=1}^\infty\left(\frac{\alpha_s}{4\pi}\right)^n {\boldsymbol Z}^{(n)}\,,\quad \widetilde {\boldsymbol  {\cal S}}_{\Tau} = {\mathbb 1} + \sum_{n=1}^\infty\left(\frac{\alpha_s}{4\pi}\right)^n\widetilde {\boldsymbol  {\cal S}}_{\Tau}^{(n)}\,,
\end{equation}
and $\widetilde {\boldsymbol  {\cal S}}_{\Tau}^{(R)}$ admits an expansion in terms of the renormalized strong coupling $\alpha_s(\mu)$ (defined here in the $\overline{\rm MS}$ scheme)
\begin{equation}
   \widetilde {\boldsymbol  {\cal S}}^{(R)}_{\Tau} = {\mathbb 1} + \sum_{n=1}^\infty\left(\frac{\alpha_s(\mu)}{4\pi}\right)^n \widetilde {\boldsymbol  {\cal S}}^{(R),\,(n)}_{\Tau}\,,
\end{equation}
where $\mu$ is the renormalization scale. The expanded form of Eq.~\eqref{eq:renorm} up to NNLO is given in Appendix~\ref{app:renorm}.

A direct consequence of Eq.~\eqref{eq:renorm} is that the renormalized soft function $\widetilde {\boldsymbol  {\cal S}}^{(R)}_{\Tau}$ satisfies the renormalization group equation (RGE)
\begin{equation}
    \mu \frac{\rmd \widetilde {\boldsymbol  {\cal S}}^{(R)}_{\Tau}}{\rmd\mu} = \frac{1}{2}\left({\boldsymbol \Gamma}_s  \widetilde {\boldsymbol  {\cal S}}^{(R)}_{\Tau} + \widetilde {\boldsymbol  {\cal S}}^{(R)}_{\Tau}  {\boldsymbol \Gamma}^\dagger_s\right)\,,
\end{equation}
where~\cite{Bell:2023yso}
\begin{equation}
    \label{eqn:Gammas-def}
    {\boldsymbol \Gamma}_s = \sum_{(ij)}{\boldsymbol T}_i\cdot {\boldsymbol T}_j \Gamma_{\rm cusp}(\alpha_s)\, \bigg[ L_{ij} + i \pi \, \lambda_{ij} \bigg] + 2 \boldsymbol{\gamma}_s(\alpha_s) \,, ~~ L_{ij}\equiv \ln(n_i\cdot n_j/2)+\ln(\mu^2 \bar{u}^2) \,.
\end{equation}
The summation in the above equation  is performed over distinct indices $i\neq j = 1, \dots , N+2$.
Furthermore,  $\bar{u}=u e^{\gamma_E}$, and $\Gamma_{\rm cusp}$ and $\boldsymbol{\gamma}_s(\alpha_s)$ are the cusp and soft anomalous dimensions, respectively. The variable $\lambda_{ij}$ is equal to $1$ if $i,j$ are both in the initial (final) state, and $0$ otherwise.

Another important property of the soft function, both at the bare and renormalized level, is that it satisfies the non-Abelian exponentiation theorem~\cite{Gatheral:1983cz,Frenkel:1984pz}, that states that it is given by the exponentials of matrices in color space
\begin{equation}\label{eq:softExp}
    \widetilde {\boldsymbol  {\cal S}}_{\Tau} = \exp\left\{\sum_{n=1}^{\infty}\left(\frac{\alpha_s}{4\pi}\right)^n \tilde{\boldsymbol s}^{(n)}_{\Tau} \right\}\,,\quad  \widetilde {\boldsymbol  {\cal S}}^{(R)}_{\Tau} = \exp\left\{\sum_{n=1}^{\infty}\left(\frac{\alpha_s(\mu)}{4\pi}\right)^n \tilde{\boldsymbol s}^{(R),(n)}_{\Tau} \right\}\,.
\end{equation}
The building blocks in the exponent of the previous equation, consist of the maximally non-Abelian subset of the color structures that contribute to a given perturbative order, sometimes called \textit{webs}. Therefore, at each new perturbative order ${\cal O}(\alpha_s^n)$, the above structure allows one to focus on the computation of the maximally non-Abelian correction and to fix the remaining contributions by exponentiation of lower-order corrections. This property is highly non-trivial in that each contribution in the exponent is a matrix in color space.

In perturbation theory, the soft function in Eq.~\eqref{eq:softExp} can be calculated in terms of eikonal QCD squared amplitudes. 
Starting from the definition in Eq.~\eqref{eq:softExp}, we can write the perturbative expansion of the bare $N$-Jettiness soft function as
\begin{equation}
\tilde{\boldsymbol  s}_{\Tau}(u) =  \sum_{n=1}^\infty\left(\frac{\alpha_s}{4\pi}\right)^n \tilde{\boldsymbol  {s}}_{\Tau}^{(n)}(u) \,,
\end{equation}
where
\begin{equation}\label{eq:softFunctionGen}
    \tilde{\boldsymbol  {s}}_{\Tau}^{(n)} (u) = \sum_{m=1}^n \int \left(\prod_{i=1}^m[\rmd k_i]\right)\,\boldsymbol{\cal M}^{(n-m)}_s(k_1,\dots,k_m)\,e^{-u\Tau(k_1,\dots,k_m)}\,.
\end{equation}
Here, the matrix element $\boldsymbol{\cal M}^{(n-m)}_s(k_1,\dots,k_X)$ denotes the $(n-m)$-loop maximally-non-Abelian contribution\footnote{In our notation, matrix elements are stripped by factors $\alpha_s^n/(4\pi)^n$ and include symmetry factors for identical particles.} to the squared amplitude describing a process of Eq.~\eqref{eq:2.1}, which can be expressed in terms of well-known eikonal factors~\cite{Catani:2019nqv,Dixon:2019lnw,Catani:2022hkb,DelDuca:2022noh,Zhu:2020ftr,Czakon:2022dwk}.
The one-emission phase-space measure is given by $[{\rm d} k] = {\rm d}^{d}k/(2 \pi)^{d-1} \delta(k^2)\theta(k^0)$ with $d = 4-2 \ep$.

It is convenient to decompose $\tilde{\boldsymbol s}_{\Tau} (u)$ according to its color structure as follows
\begin{equation}\label{eq:soft_def_col}
    \tilde{\boldsymbol s}_{\Tau} (u) = \tilde{\boldsymbol s}^{\rm dip.}_{\Tau} (u)+\tilde{\boldsymbol s}^{\rm tri.}_{\Tau} (u) + \tilde{\boldsymbol s}^{\rm quad.}_{\Tau} (u) + \dots\,,
\end{equation}
where the dipole term $\tilde{\boldsymbol s}^{\rm dip.}_{\Tau}$ is proportional to \textit{dipole} color factors describing color correlators of two hard emitters $\colT{i}\cdot \colT{j}$ (for any $i,j=1,\dots,N+2$). This term starts at ${\cal O}(\alpha_s)$. The tripole term $\tilde{\boldsymbol s}^{\rm tri.}_{\Tau}$ describes color correlators of three hard emitters, and starts at ${\cal O}(\alpha_s^2)$. Similarly, the quadruple term $\tilde{\boldsymbol s}^{\rm quad.}_{\Tau}$ describes four-legs correlators and starts at ${\cal O}(\alpha_s^3)$, and so on for higher-point color correlators.
After expanding the exponential in Eq.~\eqref{eq:softExp} in powers of $\alpha_s$, the color structures in Eq.~\eqref{eq:soft_def_col} mix due to color algebra at any given perturbative order. For instance, the iteration of the dipole term in Eq.~\eqref{eq:soft_def_col} generates tripole-like corrections, and so forth.

An important observation is that, at the cross section level, one contracts the soft function~\eqref{eq:soft_def} with the hard function (which is also a matrix in color space) for the partonic process under consideration, and evaluates the matrix element of the resulting object between color-singlet states. In doing so, some higher-point color structures in Eq.~\eqref{eq:soft_def_col} vanish at a given perturbative order by color conservation upon acting on a color-singlet state.
For instance, through second perturbative order ${\cal O}(\alpha_s^2)$ relative to the Born level, the cross section for the scattering process $p p \to F j$ (with $F$ being a color singlet and $j$ a QCD jet) only receives contributions from the dipole term $\tilde{\boldsymbol s}^{\rm dip.}_{\Tau}$, while the process  $p p \to F jj$ receives contributions from both the dipole and tripole terms in Eq.~\eqref{eq:soft_def_col}.

Through the rest of this article, we discuss the computation of the dipole and tripole terms for arbitrary $N$ (higher-point correlators only start contributing at N$^3$LO and beyond for processes with at least two jets) through NNLO.

\subsection{Computational strategy for the dipole correlator}\label{sec:strategy}
In this section we introduce a method to calculate the dipole contribution to the soft function ${\boldsymbol  s}_{\Tau}$ for a process with $N+2$ hard emitters, see Eq.~\eqref{eq:2.1}.
It is convenient to work directly at the level of the exponent in Eqs~\eqref{eq:softExp}, and consider the dipole contribution to the $n$-th order maximally non-Abelian term
\begin{equation}\label{eq:dipoledef}
    \tilde{\boldsymbol s}^{(n),\,\text{dip.}}_{\Tau} = \sum_{i\neq j}{\boldsymbol T}_i\cdot {\boldsymbol T}_j \,   \tilde{s}^{(n)}_{\Tau,\,\{i,j\}}\,
\end{equation}
where $\tilde{s}^{(n)}_{\Tau,\,\{i,j\}}$ is a $\mathbb{C}$-number and a function of $u$, $\mu$ and all the directions of the hard emitters $n_k^\mu$ ($k=1,\dots,N+2$).
We then decompose $\tilde{s}^{(n)}_{\Tau,\,\{i,j\}}$ as the sum of an \textit{inclusive} contribution $\tilde{s}^{(n)}_{\TauBar,\,\{i,j\}}$ and a correction $\Delta \tilde{s}^{(n)}_{\Tau,\,\{i,j\}}$, i.e.
\begin{equation}
    \label{eq:decomp-inclusive-exclusive}
    \tilde{s}^{(n)}_{\Tau,\,\{i,j\}}= \tilde{s}^{(n)}_{\TauBar,\,\{i,j\}} + \Delta \tilde{s}^{(n)}_{\Tau,\,\{i,j\}}\,.
\end{equation}
The inclusive term is defined by computing the dipole correction for a simplified version of the observable evaluated as a function of the \textit{total} momentum of the real radiation, namely
\begin{equation}
\label{eq2.18}
\TauBar(k_1,\dots,k_X) = \min \left\{ k_{\rm tot} \cdot n_1,\dots,
    k_{\rm tot} \cdot n_{N+2}\right\} = \Tau\left(
    k_{\rm tot} \right)\,,
\end{equation}
where 
\be
k_{\rm tot} = 
\sum_{i=1}^X k_i\,.
\ee
This definition of the inclusive soft function connects it directly to the fully-differential soft function, which has been computed to $\cO(\alpha_s^2)$ in Refs~\cite{Li:2011zp,Li:2016axz} and through three loops in Ref.~\cite{Li:2016ctv}.
This allows for the derivation of $\tilde{\boldsymbol s}_{\TauBar,\,\{i,j\}}$ up to three loops following the discussion in Section~\ref{sec:NJettiness}.

Besides its simplicity, the inclusive soft function $\tilde{\boldsymbol s}_{\TauBar,\,\{i,j\}}$ satisfies an important property that will play a key role in the calculation of the full dipole contribution to the $N$-Jettiness soft function: the 
UV divergences 
of the dipole contributions to the inclusive and full $N$-Jettiness soft functions are identical. 
This property follows from two central observations:
\begin{itemize}
    \item The inclusive soft function in Eq.~\eqref{eq:decomp-inclusive-exclusive} is used to compute the maximally non-Abelian contributions to the exponent of Eq.~\eqref{eq:softExp}. These maximally non-Abelian terms are obtained via the phase-space integration of the correlated contributions to the soft squared amplitudes, that only have support in regions of phase space in which the rapidities of all emissions in the frame of the parent $\{i,j\}$ dipole are commensurate. This implies that, in the calculation of maximally non-Abelian corrections at any given perturbative order, all the regions of the radiation phase space that give rise to UV singularities are characterized by all the soft partons being collinear to either of the two directions of the Wilson lines that define the dipole $n^\mu_i$, $n^\mu_j$.
    \item In configurations where the radiation is collinear to a given direction, the $\Tau$ observable is only sensitive to the total momentum of soft radiation. This is a direct consequence of the linearity of the observable. It follows that in the UV divergent regions of phase space the $\Tau$ observable exactly coincides with its inclusive approximation $\TauBar$. This, in particular, implies that the two variables share the same $\epsilon$ poles, and hence the difference $\Delta \tilde{s}^{(n)}_{\Tau,\,\{i,j\}}$  at any given perturbative order ${\cal O}(\alpha_s^n)$  is always finite for $\epsilon\to 0$ and it  remains  unaffected by the UV renormalization.
\end{itemize}

This property leads to a dramatic simplification in the calculation of the $\Delta \tilde{s}^{(n)}_{\Tau,\,\{i,j\}}$ function, which becomes computable using relatively standard numerical methods. Firstly, it implies that $\Delta \tilde{s}^{(n)}_{\Tau,\,\{i,j\}}$
is an end-point contribution that is independent of $u$ (or $\tvar$ in momentum space). Secondly, while $\Delta \tilde{s}^{(n)}_{\Tau,\,\{i,j\}}$ is finite for $\epsilon\to 0$, it still contains $\epsilon$ poles of IR origin at intermediate stages of the  calculation that cancel between various contributions. In this respect, we note that the complexity of the IR structure in the calculation of $\Delta \tilde{s}^{(n)}_{\Tau,\,\{i,j\}}$, and thus the complexity of the IR subtraction that is required in its numerical computation, drops by two perturbative orders, as a consequence of the fact that for a single emission, i.e.~$n=1$, the two observables $\Tau$ and $\TauBar$ are by construction identical. This means that $\Delta \tilde{s}^{(1)}_{\Tau,\,\{i,j\}}=0$, $\Delta \tilde{s}^{(2)}_{\Tau,\,\{i,j\}}$ requires a tree-level calculation, $\Delta \tilde{s}^{(3)}_{\Tau,\,\{i,j\}}$ requires a NLO calculation, and so on.
This observation will be crucial in pushing the calculation of the dipole correction to $\Tau$ to higher perturbative orders.
Finally, we  remark  that similar ideas have appeared in the context of two-loop calculations of soft and beam functions for two-leg processes (cf. Refs~\cite{Bauer:2020npd,Abreu:2022sdc,Abreu:2022zgo,Bennett:2025jli,Abreu:2026FutureJV}). The novelty of the method presented in 
this paper lies in the fact that it is applicable to processes with an arbitrary number of hard emitters and that correction $\Delta \tilde{s}^{(n)}_{\Tau,\,\{i,j\}}$ in Eq.~\eqref{eq:decomp-inclusive-exclusive} is finite at all perturbative orders, which is crucial to tame the growth of complexity at higher orders.

\subsection{Computational strategy for the tripole correlator}
The tripole color correlator contribution $\tilde{\boldsymbol s}^{\rm tri.}_{\Tau}$ in Eq.~\eqref{eq:soft_def_col} is present  for processes with four or more
hard legs and starts at second perturbative order. At NNLO it is of real-virtual nature  which, in particular, implies that at this order the inclusive variant of the $N$-Jettiness observable Eq.~\eqref{eq2.18} agrees with that of the full observable.
Nevertheless, we find that the parametrization of the real radiation used in Section~\ref{sec:NJettiness} leads to a simplified treatment of the tripole correction as well. Using this parametrization, we will describe a calculation of the tripole term in Section~\ref{sec:NJettiness-tripole}, with additional details on the renormalization provided in Appendix~\ref{sec:NJettiness-tripole-renorm}.

\section{An example: $0$-Jettiness at NNLO}
\label{sec:0Jettiness}

In this section, we illustrate the method introduced in \sec{Setup} by applying it to the case of $0$-Jettiness at NNLO. The soft function for $0$-Jettiness has been known analytically for a long time \cite{Kelley:2011ng,Hornig:2011iu,Monni:2011gb}; it coincides with the soft function for thrust \cite{Kang:2015moa}. The availability of a closed-form expression makes this observable a convenient benchmark that allows us to present the steps of the method, and describe all ingredients entering \eq{decomp-inclusive-exclusive} in a simple and transparent way. Since for $0$-Jettiness there are only two hard legs and, therefore, just one dipole, we will omit the dipole indices in this section to simplify the notation.
We begin by deriving analytically the quantity $s^{(2)}_{\TauZeroBar}$. We then demonstrate explicitly that the poles of $s^{(2)}_{\TauZeroBar}$ coincide with those of $s^{(2)}_{\TauZero}$ so that their difference is finite.

The inclusive soft function for 0-Jettiness, $s_{\TauZeroBar}$, depends only on 2 light-like directions. Without loss of generality, we choose them as the lightlike vectors aligned with the proton beam axes in a hadronic collision, i.e. $n_1=(1,0,0,1)$ and $n_2=(1,0,0,-1)$.
According to Eq.~(\ref{eq2.18}), the determination of $s_{\TauZeroBar}$ requires the calculation of eikonal matrix elements squared with 2 hard directions, i.e. $N=0$ of \eq{2.1}, with the following variable
\begin{equation}\label{eq:TauZeroBarDef}
  \TauZeroBar(k_1,k_2,\dots,k_X) = \min \left\{ n_1 \cdot \ktot ,n_2 \cdot \ktot\right\} \,,
\end{equation}
with $k_{\rm tot} = \sum \limits_{i=1}^{X} k_i$.
Since the variable in \eq{TauZeroBarDef} is only sensitive to the light-cone projections of $\ktot^\mu$, we can obtain the inclusive soft function for 0-Jettiness from the \emph{double-differential} soft function \cite{Ravindran:2006bu,Li:2016axz,Lustermans:2019cau,Billis:2019vxg}, which parametrizes the dependence of the eikonal matrix elements on the light cone components of the total momentum.
This quantity can be obtained directly from the inclusive threshold soft function \cite{Sterman:1986aj,Catani:1989ne} and it is also connected to the eikonal limit of beam functions \cite{Lustermans:2019cau,Billis:2019vxg,Mistlberger:2025lee}.
Its constants are known through three loops  from the soft expansion of color singlet production processes~\cite{Anastasiou:2014vaa,Li:2014afw}, while logarithmically enhanced contributions are known to four loops thanks to the threshold anomalous dimensions (see for example Refs~\cite{Das:2019btv,Das:2020adl,Duhr:2022cob}).
With the same notation as in \eq{softFunctionGen}, the maximally non-Abelian contribution to the double-differential soft function can be defined in perturbation theory as
\begin{align} \label{eq:DDSFdef}
    s_\text{D.D.}^{(n)} &(w_1,w_2,\eps) \\
    &=\sum_{m=1}^n \int
    \left(\prod_{i=1}^m[\rmd k_i]\right)\,{\cal M}^{(n-m)}_{s}(k_1,\dots,k_m)\,\delta\left(w_1- n_1\cdot \ktot\right)\delta\left(w_2-n_2\cdot \ktot\right)\,,\notag
\end{align}
where again we stress that we are only considering emissions from 2 hard directions, i.e. we deal with the case $N=0$ of \eq{2.1}.
Its all order structure reads \cite{Duhr:2022cob}
\beq\label{eq:DDSFbare}
    s_\text{D.D.}^\text{bare}(w_1,w_2,\eps)=  \sum_{L=1}^\infty \left(\frac{\alpha^\text{bare}_s}{(4\pi)}\right)^L (w_1 w_2)^{(-1-L \eps)} f_{\rm MNA}^{(L)}(\epsilon)\,,
\eeq
where $f_{\rm MNA}^{(L)}(\epsilon)$ is a function of $\epsilon$ and color factors, encoding the maximally non-Abelian contributions.
By fixing the value of the variable in \eq{TauZeroBarDef} we can obtain the maximally-non-Abelian correction to the inclusive 0-Jettiness soft function
\begin{align}\label{eq:stau0expr}
    s_{\TauZeroBar}(\tvar,\epsilon) &=
    \int_0^\infty \df w_1 \df w_2    \,s_\text{D.D.}^\text{bare}(w_1,w_2,\eps) \delta(\tvar - \TauZeroBar(\ktot)) \nn\\
    &= \int_0^\infty \df w_1 \df w_2  \sum_L \left(\frac{\alpha^\text{bare}_s}{(4\pi)}\right)^L (w_1 w_2)^{(-1-L \eps)} f_{\rm MNA}^{(L)}(\epsilon)\big[ 2\delta(\tvar - w_1)\theta(w_2 - w_1) \big ]\nn\\
   &= 2\sum_L \left(\frac{\alpha^\text{bare}_s}{4\pi}\right)^L \frac{\ell^{(-1-2L \epsilon)}}{L \epsilon} f_{\rm MNA}^{(L)}(\epsilon)\,.
\end{align}
The maximally non-Abelian term $f_{\rm MNA}^{(L)}$ at each perturbative order can be extracted from the result in Ref.~\cite{Duhr:2022cob} by selecting the linear term in the quadratic Casimir $C_R$ of the ${\rm SU}(3)$ representation $R$ of the Wilson lines.
In Eq.~\eqref{eq:stau0expr}, we can expand in $\epsilon$ the observable dependence using the standard distribution expansion~
\beq\label{eq:plusDexpansion}
    \tvar^{-1+a \epsilon}\mu^{-a \eps} = \frac{\delta(\tvar)}{a \epsilon} + \sum_{n=0} \frac{(a \epsilon)^n}{n!}\frac{1}{\mu}\cL_n(\tvar/\mu)\,.
\eeq
We note that at  any fixed order in $\alpha_s$ and $\epsilon$ expansion, going  from momentum to  Laplace space 
for these soft functions is straightforward, see \app{transform} for details. For this reason in this section we stay in momentum space with the understanding that any result can be straightforwardly transformed into Laplace space following the rules in \eq{laplace-rules}.
After $\alpha_s$ renormalization
\be
\left(\frac{\alpha^\text{bare}_s}{4\pi}\right)
= \left(\frac{\alpha_s(\mu)}{4 \pi}\right) \mu^{2\eps}\frac{ (4\pi)^{-\epsilon}\,e^{\eps \gamma_\eps}}{\Gamma(1-\eps)}
\left [ 1 - \left(\frac{\alpha_s(\mu)}{4 \pi}\right) \frac{\beta_0}{\eps} +
\dots \right ],
\ee
we obtain the following expression for $s_{\TauZeroBar}$ in momentum space at two loops\footnote{Our convention is to normalize the dipole coefficients by the color structure $\sum_{i\neq j}\mathbf{T}_i \cdot \mathbf{T}_j$ which, for 0-Jettiness,  gives $\sum_{i\neq j}\mathbf{T}_i \cdot \mathbf{T}_j= -2C_R$.}
\begin{align}\label{eq:stauzeroinc}
s^{(2)}_{\TauZeroBar} &=
\red{\frac{1}{\epsilon^3}}\,
\delta(\tvar)\Bigl[-\frac{11}{2}\,C_A + n_f\Bigr] \nn \\[4pt]
&
+ \red{\frac{1}{\epsilon^2}}\Bigl\{
\frac{1}{\mu}\cL_0(\tvar/\mu)\Bigl[\frac{22}{3}\,C_A - \frac{4}{3}\,n_f\Bigr]
+ \delta(\tvar)\Bigl[-\frac{5}{9}\,n_f
    + C_A\Bigl(\frac{67}{18} - \frac{\pi^2}{6}\Bigr)\Bigr]
\Bigr\} \nn \\[4pt]
&
+ \red{\frac{1}{\epsilon}}\Bigl\{
\frac{1}{\mu}\cL_0(\tvar/\mu)\Bigl[\frac{20}{9}\,n_f
    + C_A\Bigl(-\frac{134}{9} + \frac{2\pi^2}{3}\Bigr)\Bigr]
\notag \\ & \hspace{3cm} + \delta(\tvar)\Bigl[
      n_f\Bigl(-\frac{28}{27} + \frac{\pi^2}{18}\Bigr)
      + C_A\Bigl(\frac{202}{27} - \frac{11\pi^2}{36} - 7\zeta_3\Bigr)
    \Bigr]
\Bigr\} \nn \\[4pt]
&
+ \frac{1}{\mu}\cL_2(\tvar/\mu)\Bigl[-\frac{88}{3}\,C_A + \frac{16}{3}\,n_f\Bigr]
+ \frac{1}{\mu}\cL_1(\tvar/\mu)\Bigl[-\frac{80}{9}\,n_f
    + C_A\Bigl(\frac{536}{9} - \frac{8\pi^2}{3}\Bigr)\Bigr] \nn \\[4pt]
&
+ \frac{1}{\mu}\cL_0(\tvar/\mu)\Bigl[
      n_f\Bigl(\frac{112}{27} - \frac{4\pi^2}{9}\Bigr)
      + C_A\Bigl(-\frac{808}{27} + \frac{22\pi^2}{9} + 28\zeta_3\Bigr)
    \Bigr] \nn \\[4pt]
&
+ \delta(\tvar)\Bigl[
      C_A\Bigl(\frac{1214}{81} - \frac{67\pi^2}{36}
          - \frac{\pi^4}{18} - \frac{55\zeta_3}{9}\Bigr)
      + n_f\Bigl(-\frac{164}{81} + \frac{5\pi^2}{18}
          + \frac{10\zeta_3}{9}\Bigr)
    \Bigr]\,.
\end{align}
As we see, even after $\alpha_s$ renormalization, this quantity has poles in $\epsilon$. However, following the argument in \sec{strategy} we expect these poles to exactly match the corresponding ones for the 0-Jettiness soft function. At this order, the maximally non-Abelian term of the $\alpha_s$-renormalized 0-Jettiness soft function~\cite{Baranowski:2020xlp} reads
\begin{align}\label{eq:stauzerotrue}
s_{\TauZero}^{(2)} &=
\red{\frac{1}{\epsilon^3}}\,
\delta(\tvar)\Bigl[-\frac{11}{2}\,C_A + n_f\Bigr] \nn \\[4pt]
&
+ \red{\frac{1}{\epsilon^2}}\Bigl\{
\frac{1}{\mu}\cL_0(\tvar/\mu)\Bigl[\frac{22}{3}\,C_A - \frac{4}{3}\,n_f\Bigr]
+ \delta(\tvar)\Bigl[-\frac{5}{9}\,n_f
    + C_A\Bigl(\frac{67}{18} - \frac{\pi^2}{6}\Bigr)\Bigr]
\Bigr\} \nn \\[4pt]
&
+ \red{\frac{1}{\epsilon}}\Bigl\{
\frac{1}{\mu}\cL_0(\tvar/\mu)\Bigl[\frac{20}{9}\,n_f
    + C_A\Bigl(-\frac{134}{9} + \frac{2\pi^2}{3}\Bigr)\Bigr]
\notag \\ & \hspace{3cm}+ \delta(\tvar)\Bigl[
      n_f\Bigl(-\frac{28}{27} + \frac{\pi^2}{18}\Bigr)
      + C_A\Bigl(\frac{202}{27} - \frac{11\pi^2}{36} - 7\zeta_3\Bigr)
    \Bigr]
\Bigr\} \nn \\[4pt]
&
+ \frac{1}{\mu}\cL_2(\tvar/\mu)\Bigl[-\frac{88}{3}\,C_A + \frac{16}{3}\,n_f\Bigr]
+ \frac{1}{\mu}\cL_1(\tvar/\mu)\Bigl[-\frac{80}{9}\,n_f
    + C_A\Bigl(\frac{536}{9} - \frac{8\pi^2}{3}\Bigr)\Bigr] \nn \\[4pt]
&
+ \frac{1}{\mu}\cL_0(\tvar/\mu)\Bigl[
      n_f\Bigl(\frac{112}{27} - \frac{4\pi^2}{9}\Bigr)
      + C_A\Bigl(-\frac{808}{27} + \frac{22\pi^2}{9} + 28\zeta_3\Bigr)
    \Bigr] \nn \\[4pt]
&
+ \delta(\tvar)\Bigl[
      C_A\Bigl(\frac{1070}{81} + \frac{335\pi^2}{108}
          - \frac{11\pi^4}{45} - \frac{319\zeta_3}{9}\Bigr)
      + n_f\Bigl(-\frac{20}{81} - \frac{37\pi^2}{54}
          + \frac{58\zeta_3}{9}\Bigr)
    \Bigr]\,.
\end{align}
Comparing \eq{stauzeroinc} and \eq{stauzerotrue},
we observe that 
$\ep$-poles 
exactly match. 
The finite difference between the two soft functions, which we denote as $\Delta s^{(2)}_{\TauZero}$ following the notation of \eq{decomp-inclusive-exclusive}, reads
\begin{equation}\label{eq:deltaTau0}
    \Delta s^{(2)}_{\TauZero} = \delta(\tvar)\Bigl[
      C_A\Bigl(-\frac{16}{9} + \frac{134\pi^2}{27} - \frac{17\pi^4}{90} - \frac{88\zeta_3}{3}\Bigr)
      + n_f\Bigl(\frac{16}{9} - \frac{26\pi^2}{27} + \frac{16\zeta_3}{3}\Bigr)
    \Bigr]\,.
\end{equation}
\sloppy
\section{N-Jettiness at NNLO}
\label{sec:NJettiness}
We now extend the discussion to the case of generic $N$. We start with the analysis of  the inclusive contribution, and then continue with the non-inclusive one.

\subsection{The dipole correlator: calculation of $s^{(2)}_{\TauBar,\,\{i,j\}}$}
\label{sec:NJ-inclusive}
To compute the inclusive $N$-Jettiness soft function, we need to integrate the eikonal matrix elements at a fixed value of the inclusive Jettiness
defined in Eq.~(\ref{eq2.18}), 
which we show here one more time
\begin{equation}
\label{eq:tauNincl}
    \TauBar(\ktot) = \min \left\{ n_1 \cdot \ktot,n_2 \cdot \ktot,\dots,n_{N+2} \cdot \ktot\right\}\,,
    \;\;\;
    \ktot = k_1 + k_2.
\end{equation}
For a given dipole $(i,j)$ we define the variables
\begin{equation}
\label{eq:def-LCD-ijDIP}
w_i \equiv n_i\!\cdot\!\ktot,\qquad
w_j \equiv n_j\!\cdot\!\ktot,\qquad
x \equiv \frac{(n_i\!\cdot\!n_j)\,\ktot^2}
{2\,(n_i\!\cdot\!\ktot)(n_j\!\cdot\!\ktot)}\,.
\end{equation}
and for any other hard direction $A\neq i,j$ we introduce the purely geometric coefficients
\begin{equation}
c_{Ai}\equiv \frac{n_A\!\cdot\!n_i}{n_i\!\cdot\!n_j},
\qquad
c_{Aj}\equiv \frac{n_A\!\cdot\!n_j}{n_i\!\cdot\!n_j}.
\end{equation}
Parametrizing the hard directions by angles, $n_r^\mu=(1,\vec n_r)$ with $|\vec n_r|=1$, and
$\cos\Theta_{rs}\equiv \vec n_r\!\cdot\!\vec n_s$, one has
\begin{equation}
\label{eqn:def-LCD-cAi-cAj}
n_r\!\cdot\!n_s = 1-\cos\Theta_{rs}
\quad\Longrightarrow\quad
c_{Ai}=\frac{1-\cos\Theta_{Ai}}{1-\cos\Theta_{ij}},
\qquad
c_{Aj}=\frac{1-\cos\Theta_{Aj}}{1-\cos\Theta_{ij}}.
\end{equation}
Using these definitions in \eq{tauNincl} we find
\begin{align}\label{eq:nA-dot-ktot}
n_A\!\cdot\!\ktot
&=
c_{Aj}\,w_i + c_{Ai}\,w_j
-2\sqrt{c_{Ai}c_{Aj}\,w_i w_j}\,\sqrt{1-x}\,\cos(\phi-\phi_A) \notag \\ &\equiv
w_A(w_i,w_j,x,\phi;\vec{\Theta}) \,,
\end{align}
where $\phi$ is the azimuthal angle of $\ktot^\perp$, and $\phi_A$ is the fixed azimuthal angle  of the transverse projection of $n_A$, both defined in the plane orthogonal to the dipole $(n_i,n_j)$.
In this way we rewrite the inclusive $N$-Jettiness constraint as a function of $(w_i,w_j,x,\phi)$ and of the set of hard-direction angles $\vec\Theta$,
\begin{equation}
\TauBar(\ktot)
=\min\Bigl\{w_i,\;w_j,\;\{\,w_A(w_i,w_j,x,\phi;\vec{\Theta})\,\}_{A\neq i,j}
\Bigr\} \equiv {\TauBar}(w_i,w_j,x,\phi;\vec{\Theta}),
\end{equation}
with $n_A\!\cdot\!\ktot$ given in Eq.~\eqref{eq:nA-dot-ktot}.
Therefore, we can generalize the reasoning in \sec{0Jettiness}, and rewrite the maximally non-abelian term of the inclusive $N$-Jettiness soft function as a projection of the \emph{fully differential} soft function~\cite{Li:2011zp}, which is analytically known through three loops~\cite{Li:2016ctv}.
Concretely, at order $\cO(\alpha_s^L)$ in perturbation theory, the maximally-non-Abelian contribution to the fully differential soft function is defined as
\begin{align}\label{eq:njet-fd-def}
s^{(L)}_{\text{F.D.}}(w_i,w_j,x)&\equiv
\sum_{m=1}^{L}\int\!\left(\prod_{r=1}^{m}[\rmd k_r]\right)
{\cal M}^{(L-m)}_{s,\{i,j\}}(k_1,\dots,k_m) \,\nn\\
&\qquad\times\,
\delta\!\left(w_i-n_i\!\cdot\!\ktot\right)\,
\delta\!\left(w_j-n_j\!\cdot\!\ktot\right)\,
\delta\!\left(x-\frac{(n_i\!\cdot\!n_j)\,\ktot^2}
{2\,(n_i\!\cdot\!\ktot)(n_j\!\cdot\!\ktot)}\right)\,.
\end{align}
Note that the fully differential soft function is also related to the double differential soft one of \eq{DDSFdef} via the sum rule
\beq\label{eq:SDDfromSFD}
\int_0^1 \df x\, s_{\text{F.D.}}(w_i,w_j,x) = s_{\text{D.D.}}(w_i,w_j)\,.
\eeq
Its relation to  the maximally non-abelian term of the inclusive $N$-Jettiness soft function at $\cO(\alpha^L_s)$, $s^{(L)}_{\TauBar,\{i,j\}}$ is described 
by the following equation
\begin{align}\label{eq:njet-sincl}
s^{(L)}_{\TauBar,\{i,j\}}&(\tvar;\vec{\Theta})= \\
&\int  [{\rm d}\Omega^\perp]_\eps \int_0^1\!\df x\!\int_0^\infty\!\df w_i\df w_j\,
s^{(L)}_{\text{F.D.}}(w_i,w_j,x)\,
\delta\!\left(\tvar-{\TauBar}(w_i,w_j,x,\phi;\vec{\Theta})\right)\,.\notag
\end{align}
where  the normalized solid angle in $d-2$ dimensions $[{\rm d}\Omega^\perp]_\eps$ reads 
\begin{equation}\label{eq:domega_perp}
[{\rm d}\Omega^\perp]_\eps \equiv \frac{{\rm d}\Omega^{(2-2\eps)}}{\Omega^{(2-2\ep)}}\,.
\end{equation}
The bare fully differential soft function takes the form \cite{Li:2016axz,Duhr:2022cob}
\beq
s_{\text{F.D.}}(w_i,w_j,x,\ep) =  \sum_{L=1}^\infty \left(\frac{\alpha^\text{bare}_s}{(4\pi)}\right)^L (w_i w_j)^{(-1-L \eps)} \left(\frac{n_i \cdot n_j}{2}\right)^{L \eps} f_{\rm MNA}^{(L)}(x,\epsilon)\,,
\eeq
where, order by order in $\epsilon$, $f_{\rm MNA}^{(L)}(x,\epsilon)$ can be expanded in terms of regular functions of $x$ and distributions at $x=0$
\beq
f_{\rm MNA}^{(L)}(x,\epsilon)=
\sum_k \eps^k \left[f^{(L,k)}_{\text{reg}}(x)
+ \sum_{m=0}^{2L -1} \cL_m(x) f^{(L,k)}_{\cL_m}
+  \delta(x) f^{(L,k)}_{\delta}\right]\,.
\eeq
We note that the sum rule \eq{SDDfromSFD} implies a sum rule for $f$ itself
\be\label{eq:sumrulef}
   \int_0^1 \df x f_{\rm MNA}^{(L)}(x,\epsilon) = f_{\rm MNA}^{(L)}(\epsilon)\,,
\ee
where $f_{\rm MNA}^{(L)}(\epsilon)$ was defined in \eq{DDSFbare}. As for the function $f_{\rm MNA}^{(L)}$, the quantity $f_{\rm MNA}^{(L)}(x,\epsilon)$  can be extracted from the result in Ref.~\cite{Li:2016axz} by selecting the linear term in the quadratic Casimir $C_R$ at each perturbative order.

We now specialize to the bare case at NNLO ($L=2$) and show how to obtain a practical integral representation for $s^{(2)}_{\TauBar,\{i,j\}}(\tvar;\vec{\Theta})$.
Since $\TauBar$ is a  homogeneous function of degree one w.r.t. variables $(w_i,w_j)$, we introduce an overall energy scale $E$ and a momentum fraction $z\in[0,1]$ such that $w_i=E\,z$ and $w_j=E(1-z)$, and factor out $E$ from the measurement function to define the rescaled observable $t$,
\beq\label{eq:t-def}
t_{\{i,j\}}(z,x,\phi;\vec{\Theta}) = \min\Bigl\{z,\; 1-z,\; \bigl\{w_A(z,1-z,x,\phi;\vec{\Theta})\bigr\}_{A\neq i,j}\Bigr\}\,.
\eeq
Using this representation to factor  out the dependence on $\tvar$, we obtain 
\begin{align}
    \label{eq:STauNTauOut}
    &s^{(2)}_{\TauBar,\{i,j\}}(\tvar,\eps) \\
    &{} = \left(\frac{n_i \cdot n_j}{2}\right)^{2 \eps} \tvar^{-1-4\epsilon} 
    \int [{\rm d}\Omega^\perp]_\eps \int_0^1\!\df x\int_0^1\!\df z\;
    \bigl[z(1\!-\!z)\bigr]^{-1-2\epsilon}\;
    t^{4\epsilon}_{\{i,j\}}(z,x,\phi;\vec{\Theta})\; f^{(2)}(x,\epsilon)\,.\nonumber
\end{align}
While we can easily expand in $\epsilon$ the $\tvar$-dependent pre-factor
using \eq{plusDexpansion}, we cannot evaluate the integral in \eq{STauNTauOut} order by order in $\eps$, as it contains collinear poles at the endpoints of the $z$ integration.\footnote{Note that we cannot expand the factor $\bigl[z(1-z)\bigr]^{-1-2\epsilon}$ in distributions in $z$ (or $1-z$), because $t^{4\epsilon}_{{i,j}}(z,x,\phi;\vec{\Theta})$ is in general not a regular function of $z$ as $z\to 0,1$.}
The physical origin of these endpoint singularities is straightforward to understand: the limits $z\to 0$ and $z\to 1$ correspond to radiation becoming collinear to $n_i$ and $n_j$, respectively. In these strict collinear limits, the measurement must be insensitive to the geometry of the full $N$-leg hard configuration, so the singular behavior should be related to the one already obtained for inclusive 0-Jettiness.
To make this relation manifest and isolate this endpoint collinear singularity analytically in a $N$-leg geometry-independent way, we split $t$ against the minimum appropriate to the $\TauZeroBar$ case,
\beq\label{eq:t0-def}
t_0(z) \equiv \min\{z,1-z\}\,.
\eeq
Since extra hard directions can only lower the minimum, one has $t_{\{i,j\}}\le t_0$ pointwise. Writing
\beq
t_{\{i,j\}}^{4\epsilon}=t_0^{4\epsilon}+\bigl[t_{\{i,j\}}^{4\epsilon}-t_0^{4\epsilon}\bigr] \,,
\eeq
we obtain 
\begin{align}\label{eq:STauNTauOutsplit}
    s^{(2)}_{\TauBar,\{i,j\}} (\tvar,\eps) = {} & \left(\frac{n_i \cdot n_j}{2}\right)^{2 \eps} \Bigg[s^{(2)}_{\TauZeroBar}(\tvar,\eps) \\
    &+ \tvar^{-1-4\epsilon}\int [{\rm d}\Omega^\perp]_\eps\int_0^1\!\df x\int_0^1\!\df z\;
    \frac{\Bigl[t_{\{i,j\}}^{4\epsilon}-t_0^{4\epsilon}\Bigr](z,x,\phi;\vec{\Theta})}{\bigl[z(1\!-\!z)\bigr]^{1+2\epsilon}}\; f^{(2)}(x,\epsilon)\Bigg]\,,\notag
\end{align}
where, in order to identify the $t_0^{4\epsilon}$ dependent term with $s^{(2)}_{\TauZeroBar}(\tvar,\eps)$, we used the fact that $t_0$ is independent of $x$, together with the sum rule in \eq{sumrulef}.
The entire dependence on the 
directions of $N$ jets is now encoded in the integral containing $\bigl(t^{4\epsilon}-t_0^{4\epsilon}\bigr)$.
Crucially, this difference vanishes whenever the radiation is closer to either leg of the dipole, and in particular in the collinear regions $z\to 0,1$.
As a result, the $N$-leg geometry-dependent term in \eq{STauNTauOutsplit} is free of collinear divergences and can be evaluated numerically after expanding the integrand in $\eps$ to the desired order.

\subsection{The dipole correlator: calculation of $\Delta \tilde{s}^{(2)}_{\TauN{N},\,\{i,j\}}$}
\label{sec:NJettiness-num-deltaij}
In this section, we show how to obtain the difference of the two Jettinesses $\Delta \tilde{s}^{(2)}_{\TauN{N},\,\{i,j\}}$ numerically.  This term receives contributions only from configurations with the emission of two real soft partons, since by construction it vanishes for a single soft emission. Consequently, the soft integral relevant for $\Delta \tilde{s}^{(2)}_{\TauN{N},\,\{i,j\}}$ takes the form
\begin{equation}\label{eq:deltas-tauzero-rr}
\Delta \tilde{s}^{(2)}_{\TauN{N},\,\{i,j\}} \equiv \int [\rmd k_1]\,[\rmd k_2]\, {\cal M}_{s,\{i,j\}}^{(0)}(k_1,k_2)
\left(e^{-u\, \TauN{N}(k_1,k_2)} - e^{-u\, \TauN{N}(k_1+k_2)}\right)\,,
\end{equation}
where ${\cal M}_{s,\{i,j\}}^{(0)}(k_1,k_2)$ denotes the maximally non-Abelian part of the tree-level double-soft squared amplitude~\cite{Catani:1999ss} associated with the dipole $\{i,j\}$. We provide its expression in Eq.~\eqref{eq:Ms-RR-nab}.
Although the soft matrix element squared ${\cal M}_{s,\{i,j\}}^{(0)}(k_1,k_2)$ depends solely on the dipole $\{i,j\}$, the observable $\TauN{N}$ involves all hard directions. Consequently, the integral can be computed numerically after fixing the kinematic configuration of the hard legs, 
i.e. once the set of directions $\{n_k\}_{k=1}^N$ is specified.
In any singular configuration of ${\cal M}_{s,\{i,j\}}^{(0)}(k_1,k_2)$, the observable satisfies $\TauN{N}(k_1,k_2)\to\TauN{N}(k_1+k_2)$, causing the expression in parentheses to vanish. Therefore, the integral is free of infrared singularities and can be immediately  evaluated  in four space-time dimensions.

To make the cancellation of divergences manifest, it is convenient to perform some integrations analytically. We parametrize the $d$-dimensional double-soft phase space in terms of energies and angles of the emitted gluons, defined in the partonic centre-of-mass frame, as
\begin{equation}
\label{eqn:def-para-energy-angle}
[\rmd k_1]\,[\rmd k_2] = E_1^{1-2\epsilon} \rmd E_1 \; [\rmd\Omega_1]_\epsilon E_2^{1-2\epsilon} \rmd E_2 \; [\rmd\Omega_2]_\epsilon \,,
\end{equation}
where $[\rmd \Omega]_\epsilon  = \rmd \Omega^{(3-2\epsilon)} / (2(2\pi)^{3-2\epsilon})$.
Given the symmetry of the integrand under the exchange of $1 \leftrightarrow 2$, we can select a specific energy ordering of two soft momenta,
$E_2 < E_1$, and supply a factor of two.  Then, we rescale  $E_2 = \xi E_1$, with $0 \leq\xi \leq 1$. Under this transformation, the matrix element ${\cal M}_{s,\{i,j\}}^{(0)}(k_1,k_2)$ becomes 
\begin{equation}
{\cal M}_{s,\{i,j\}}^{(0)}(k_1,k_2) = \frac{1}{E_1^4}{\cal M}_{s,\{i,j\}}^{(0)}(\hat{k}_1,\xi \hat{k}_2)\,,
\end{equation}
where $\hat{k}_{1,2} = (1,\hat{\vec{k}}_{1,2})$ and $\hat{\vec{k}}_{1,2}$ denote the unit vectors along the spatial momenta $\vec{k}_{1,2}$.
Similarly, the observable rescales as
\begin{equation}
\TauN{N}(k_1,k_2) = E_1 \TauN{N}(\hat{k}_1,\xi \hat{k}_2), \quad \TauN{N}(k_1+k_2) = E_1 \TauN{N}(\hat{k}_1 + \xi \hat{k}_2) \,.
\end{equation}
This leads to
\begin{align}
\Delta \tilde{s}^{(2)}_{\TauN{N},\,\{i,j\}}  =
2\int [ {\rm d} \Omega_{12}]_\epsilon \; &
\int_{0}^{1}
 {\rm d} \xi \; \xi^{1-2\ep} \;
\int_{0}^{\infty}  \frac{ {\rm d} E_1 }{E_1^{1+4\ep}} \;
{\cal M}_{s,\{i,j\}}^{(0)}(\hat{k}_1,\xi \hat{k}_2) \notag \\ &~\times \left (
e^{-u\, E_1 \TauN{N}(\hat{k}_1,\xi \hat{k}_2)}
-e^{-u \, E_1 \TauN{N}(\hat{k}_1+\xi \hat{k}_2)}
\right ),
\end{align}
where we introduced the short-hand notation $[{\rm d} \Omega_{12}]_\epsilon = [{\rm d} \Omega_1]_\epsilon [{\rm d} \Omega_2]_\epsilon$.
The integration over $E_1$ can be carried out analytically, yielding
\begin{align}
\Delta \tilde{s}^{(2)}_{\TauN{N},\,\{i,j\}}  =
\frac{N_\ep}{2\epsilon}
\int [ {\rm d} \Omega_{12} ]_\epsilon
\int \limits_{0}^{1}
\frac{{\rm d} \xi}{ \xi^{-1+2\ep}}
{\cal M}_{s,\{i,j\}}^{(0)}(\hat{k}_1,\xi \hat{k}_2) \; 
\left[
\TauZero^{4\epsilon}(k_{12,\xi})
-
\Tau^{4\epsilon}(\hat{k}_1,\xi \hat{k}_2)\right]\,,
\label{eq:deltas-tauzero-rr-w1int}
\end{align}
with $N_\ep = u^{4\epsilon} \,\Gamma(1-4\epsilon)$ 
and $k_{12,\xi} = k_1 + \xi k_2$.

Provided that the integrals over $\xi$ and the angular variables $\Omega_1$ and $\Omega_2$ do not generate further singularities, we may expand Eq.~\eqref{eq:deltas-tauzero-rr-w1int} in powers of $\ep$ at the integrand level. Keeping only the leading term in the $\ep$-expansion, we obtain
\begin{equation}
\Delta \tilde{s}^{(2)}_{\Tau,\,\{i,j\}} =
2 \,\int [ {\rm d} \Omega_{12} ]_{\epsilon=0}
\int \limits_{0}^{1} {\rm d} \xi
\;
\xi  \;
{\cal M}_{s,\{i,j\}}^{(0)}(\hat{k}_1,\xi \hat{k}_2)
\ln \left ( \frac{\Tau(\hat{k}_1+\xi \hat{k}_2)}{\Tau(\hat{k}_1,\xi \hat{k}_2)}
\right )\,.
\label{eq:deltas-tauzero-rr-final}
\end{equation}
We can verify the assumption about  finiteness  of  $\Delta \tilde{s}^{(2)}_{\Tau,\,\{i,j\}} $
a posteriori,  by analyzing the above integral.
All potential divergences should arise from the singular behavior of ${\cal M}_{s,\{i,j\}}^{(0)}(\hat{k}_1,\xi \hat{k}_2)$ in the following limits: the soft limit $\xi \to 0$, the collinear configuration $\hat{\vec{k}}_1 \parallel \hat{\vec{k}}_2$, and the triple-collinear limits $\hat{\vec{k}}_1 \parallel \hat{\vec{k}}_2 \parallel \vec{n}_i$ or $\hat{\vec{k}}_1 \parallel \hat{\vec{k}}_2 \parallel \vec{n}_j$. In each of these cases, the observable satisfies $\TauZero(\hat{k}_1,\xi \hat{k}_2)\to\TauZero(\hat{k}_1+\xi \hat{k}_2)$, so that the logarithm in Eq.~\eqref{eq:deltas-tauzero-rr-final} vanishes and the corresponding singularity is regulated. The five-dimensional integral in Eq.~\eqref{eq:deltas-tauzero-rr-final} is therefore finite without additional regulator, and can be evaluated numerically using standard Monte Carlo integration methods. 
Finally, given that $\Delta \tilde{s}^{(2)}_{\Tau,\,\{i,j\}}$ does not depend on the Laplace variable $u$, its inverse Laplace transform to momentum space amounts simply to
\begin{equation}
    \Delta \tilde{s}^{(2)}_{\TauN{N},\,\{i,j\}}  \to  \Delta \tilde{s}^{(2)}_{\TauN{N},\,\{i,j\}} \,\delta(\tvar) \equiv \Delta {s}^{(2)}_{\TauN{N},\,\{i,j\}} \,\delta(\tvar).
\end{equation}

Before concluding this section, we briefly comment on the extension of the method outlined here to N$^3$LO. Given that the NNLO correction $\Delta \tilde{s}^{(2)}_{\Tau,\,\{i,j\}}$ computed above requires the computation of tree-level (double-real) corrections, and it is finite, we expect that the computation of the $\Delta \tilde{s}^{(3)}_{\Tau,\,\{i,j\}}$ at N$^3$LO can be organized in such a way that it only requires NLO-like subtractions. Ongoing investigations seem to confirm this expectation~\cite{Buonocore:2026FutureJettiness}.

\subsection{The tripole correlator}
\label{sec:NJettiness-tripole}
Having obtained the dipole contribution to the $N$-Jettiness soft function, we will discuss the tripole contribution, which is the last missing piece at NNLO.
Following the normalization conventions of Eq.~\eqref{eq:softExp}, we write the perturbative expansion of the Laplace transform of the tripole correction as
\begin{equation}
    \tilde{\boldsymbol s}^{\rm tri.}_{\Tau} (u)  = \sum_{n=2}^\infty \left(\frac{\alpha_s}{4\pi}\right)^n\,\tilde{\boldsymbol s}^{\rm tri.,(n)}_{\Tau} (u)\,.
\end{equation}
The series starts at the second order, $n=2$, where the contribution reads
\begin{align}
    \label{eqn:def-S-trip}
    \tilde{\boldsymbol s}^{\rm tri.,(2)}_{\Tau} (u)
    = \frac{N_\ep}{\ep} \sum_{(ijk)} {\boldsymbol F}_{ijk}  \, \kappa_{jk} \, I_{\{ij\},k}(u), \quad
    N_\ep = \frac{2^{3-\ep} \pi e^{2\gamma_E \ep} \, \Gamma(1-\ep)\Gamma(1+\ep)}{\Gamma(1-2\ep)} \,.
\end{align}
In Eq.~\eqref{eqn:def-S-trip} the sum is restricted to distinct summation indices $i\neq j \neq k$, ${\boldsymbol F}_{ijk}$ and $\kappa_{jk}$ are given by
\begin{align}
    \label{eqn:def-Flmn}
    {\boldsymbol F}_{ijk}=f_{abc}\colT{i}^a\colT{j}^b\colT{k}^c\,,
\end{align}
and
\begin{align}
    \label{eqn:def-kappa}
    \kappa_{ij} = \lambda_{ij} - \lambda_{i,\text{out}} - \lambda_{j,\text{out}} \,,
\end{align}
which evaluates to $+1$ if both emitters $i,j$ are incoming and $-1$ otherwise.
Moreover, we have introduced the integral
\begin{align}
    \label{eqn:def-S-trip-int}
    I_{\{ij\},k}(u) = \int_{0}^{\infty} \mathrm{d} \tvar \, e^{-u \tvar} \int [ {\rm d} q] \;  \frac{2 (2\pi)^{3-2\eps}}{\Omega^{(2-2\eps)}} \delta(\tvar-\Tau(q)) \mathcal{M}^{(1)}_{s,\{\{ij\},k\}}(q) \, ,
\end{align}
In the above equation, $\mathcal{M}^{(1)}_{s,\{\{ij\},k\}}(q)$ denotes the color-stripped tripole contribution, c.f. Eq.~\eqref{eq:Ms-RV-tripole}, which develops singularities when $q$ becomes collinear to either $n_i$ or $n_j$.
Therefore, in each triple-correlator contribution $\{\{ij\},k\}$, we parametrize all momenta using the Sudakov decomposition with respect to the light-cone directions $\{n_i,n_j\}$ as in Section~\ref{sec:NJettiness}. In particular, for a light-like  four-vector $q$, we write  
\begin{align}
    q = \frac{w_j}{n_i\cdot n_j} n_i +  \frac{w_i}{n_i\cdot n_j} n_j + \sqrt{\frac{2 w_i w_j}{n_i\cdot n_j}} e_\perp\,,
\end{align}
where  $e_\perp^2=-1$. 
With this parametrization, the single-particle phase space becomes
\begin{align}\label{eq:ps-one}
    \frac{{\rm d}^d q}{(2\pi)^{d-1}} \, \delta^+(q^2) & = [{\rm d} q ] = \frac1{n_i\cdot n_j} (w_i w_j)^{-\ep} \left( \frac{n_i\cdot n_j}{2}\right)^{\ep} {\rm d} w_i \, {\rm d} w_j \, \frac{{\rm d}\Omega^{(2-2\eps)}}{2(2\pi)^{3-2\eps}}\,.
\end{align}
Following the notation introduced in Eq.~\eqref{eq:nA-dot-ktot}, we denote with
\begin{align}\label{eq:wA}
w_l(w_i,w_j,0,\phi_{q}) &= q\cdot n_l \\ \notag
& = w_i \frac{n_l\cdot n_j}{n_i\cdot n_j} + w_j \frac{n_l\cdot n_i}{n_i\cdot n_j} - \frac{2}{(n_i\cdot n_j)}\sqrt{w_i w_j (n_l\cdot n_i) (n_l\cdot n_j)} ~ \cos\phi_{ql}\,,
\end{align}
all the scalar products of $q$ with any direction $n_l$ different from the directions $\{n_i,n_j\}$ employed in the Sudakov decomposition. In the above equation we have defined $\phi_{ql}=\phi_q-\phi_l$, and we have omitted in the arguments of $w_l$ the explicit dependence on $\vec{\Theta}$ to ease the notation.
In this parametrization, the inclusive $N$-Jettiness observable reads
\begin{align}\label{eq:tauN-tripole}
\Tau(q) = \text{min}\!\left\{w_i,w_j,\{w_A(w_i,w_j,0,\phi_{q})\}_{A\ne i,j}\right\} \equiv \Tau(w_i,w_j,\phi_q;w_{A\ne i,j})\,.
\end{align}
The  integral $I_{\{ij\},k}$ becomes
\begin{align}
    \label{eqn:tripole-Ilmn-wlwmRn}
     I_{\{ij\},k}(u)  ={} & \left(\frac{(n_i\cdot n_j)(n_j\cdot n_k)}{2}\right)^{\ep} \int_{0}^{\infty} \mathrm{d} \tvar \, e^{-u\tvar}
  \int \textrm{d}w_i \textrm{d} w_j \, [{\rm d}\Omega^\perp]_\eps \notag \\
  & ~~~  \times \frac{ (w_i w_j)^{-1 -\ep}   }{\left[ w_j \, w_k(w_i,w_j,0,\phi_q) \right]^\ep} \, \delta\left(\tvar - \Tau(w_i,w_j,\phi_q;w_{A\ne i,j})\right) \,,
\end{align}
with $[{\rm d}\Omega^\perp]_\eps$ defined in \eq{domega_perp}. 

We emphasize that in this parametrization
the two collinear singularities correspond to $w_i \to 0$ and $w_j \to 0$, respectively.
We disentangle them by splitting the integration domain into two sectors by means of the decomposition $1=\theta(w_i-w_j)+\theta(w_j-w_i)$. Then, in each sector, we factor out the larger light-cone component by changing  variables $\{w_i=y,w_j=y\xi\}$ and $\{w_j=y\xi,w_i=y\}$ in the first (plus) and second sectors (minus), respectively. Using the homogeneity of $w_k$ and $\Tau$ (c.f.~Eq.~\eqref{eq:wA} and Eq.~\eqref{eq:tauN-tripole}), we obtain
\begin{align}
     I_{\{ij\},k}^{+}(u) ={} & \left(\frac{(n_i\cdot n_j)(n_j\cdot n_k)}{2}\right)^{\ep}
  \int_{0}^{\infty} \mathrm{d} \tvar \, e^{-u\tvar} \int_{0}^{\infty}  
  \textrm{d} y \, \notag \\ & \hspace{5cm} \times
  \int_{0}^{1} 
  \textrm{d}\xi \,  [{\rm d}\Omega^\perp]_\eps  \frac{ y^{-1-4\eps}  \xi^{-1 - 2\ep}   }{\, w_k^\eps(1,\xi,0,\phi_q)}  \nonumber \, \delta\left(\tvar - y \tau_{N,\{ij\}}^{+}\right) \,, \\
      I_{\{ij\},k}^{-}(u)  ={} & \left(\frac{(n_i\cdot n_j)(n_j\cdot n_k)}{2}\right)^{\ep}
 \int_{0}^{\infty} \mathrm{d} \tvar \, e^{-u\tvar} \int_{0}^{\infty} \textrm{d} y \notag \\ & \hspace{5cm} \times \int _{0}^{1} \textrm{d}\xi \,  [{\rm d}\Omega^\perp]_\eps   \frac{ y^{-1-4\eps}  \xi^{-1 - \ep}    }{ w_k^\eps(\xi,1,0,\phi_q)}  \nonumber \, \delta\left(\tvar - y \tau_{N,\{ij\}}^{-}\right) \,,
\end{align}
where we have introduced the functions $\tau_{N,\{ij\}}^{+} (\xi,\phi_{q}) \equiv \Tau(1,\xi,\phi_q,\{w_A(1,\xi,0,\phi_q)\}_{A\ne i,j})$ and $\tau_{N,\{ij\}}^{-} (\xi,\phi_{q}) \equiv \Tau(\xi,1,\phi_q,\{w_A(\xi,1,0,\phi_q)\}_{A\ne i,j})$ that depend on the directions $n_i$, $n_j$, but not on the variable $y$.

We remove  the delta function by integrating over $y$, and use Eq.~\eqref{eqn:LT-tauaep} to perform the Laplace integral over $\tvar$. We find
\begin{align}
\int_{0}^{\infty} \mathrm{d} \tvar \, e^{-u\tvar} \int_{0}^{\infty} \textrm{d} y \, y^{-1-4\ep} \delta\left(\tvar - y \, \tau_{N,\{ij\}}^{\pm} \right) &= \frac{\bar{u}^{4\ep} e^{-4\gamma_E \ep} \Gamma(1-4\ep)}{-4\ep} \left(\tau_{N,\{ij\}}^{\pm}\right)^{4\eps} \notag \\  & = \frac{\bar{u}^{4\ep} e^{-4\gamma_E \ep} \Gamma(1-4\ep)}{-4\ep} \xi^{4\eps} \left(t_{N,\{ij\}}^{\pm}\right)^{4\eps}\,,
\end{align}
where, in the last step, we have factored out the leading behavior of $\tau_{N,\{ij\}}^{\pm}$ in the singular limit $\xi\to0$. The functions $t_{N,\{ij\}}^{\pm}$ represent, therefore, the explicit parametrization of the rescaled inclusive $N$-Jettiness observable in the two sectors
\begin{align}
    \label{eqn:ti-tj-NLO-def}
    t_{N,\{ij\}}^{+}(\xi,\phi_{q}) \equiv \frac{\Tau(q/y)}{\xi} &= \frac{\text{min}\!\left\{\xi, \{w_A(1,\xi,0,\phi_q)\}_{A\ne i,j}\right\}}{\xi} \,,\\
    t_{N,\{ij\}}^{-}(\xi,\phi_{q}) \equiv \frac{\Tau(q/y)}{\xi} &=  \frac{\text{min}\!\left\{\xi, \{w_A(\xi,1,0,\phi_q)\}_{A\ne i,j}\right\}}{\xi} \,,
\end{align}
and, by construction, satisfy the following equation
\begin{equation}
    t_{N,\{ij\}}^{+}(0,\phi_{q}) = t_{N,\{ij\}}^{-}(0,\phi_{q}) = 1\,.
\end{equation}
We obtain
\begin{align}\label{eq:Iijk_parametrization}
    I_{\{ij\},k}(u) = -&\frac{\bar{u}^{4\ep} e^{-4\gamma_E \ep} \Gamma(1-4\ep)}{4\ep} \left(\frac{(n_i\cdot n_j)(n_j\cdot n_k)}{2}\right)^{\ep} \int_{0}^{1} {\rm d} \xi  \, \int [{\rm d} \Omega^\perp]_\epsilon \notag \\ &\times \left[\xi^{-1+2\eps}\frac{\left(t_{N,\{ij\}}^{+}(\xi,\phi_{q})\right)^{4\eps}}{w_k^\eps(1,\xi,0,\phi_q)} + \xi^{-1+3\eps}\frac{\left(t_{N,\{ij\}}^{-}(\xi,\phi_{q})\right)^{4\eps}}{w_k^\eps(\xi,1,0,\phi_q)} \right]\,.
\end{align}

In the explicit parametrizations of Eq.~\eqref{eq:Iijk_parametrization}, the one-emission collinear singularities are mapped to the point $\xi=0$ and are encapsulated in the singular factors $\xi^{-1+a\eps}$, $a =2,3$.

With this parametrization, we can easily perform the renormalization of the tripole correlator, which we report in detail in Appendix~\ref{sec:NJettiness-tripole-renorm}. After the cancellation of the UV poles against the counter-terms, we write the finite remainder of the renormalized tripole correction as
\begin{align}
    \tilde{\boldsymbol s}^{\rm tri.,(2),(R)}_{\Tau} (u)
=    \sum_{(ijk)} {\boldsymbol F}_{ijk}  \kappa_{jk}
    \tilde{ s}^{\rm tri.,(2),(R)}_{\Tau,\{i,j,k\}}(u) \; ,
\end{align}
where
\begin{align}\label{eq:triple_renorm}
    \tilde{ s}^{\rm tri.,(2),(R)}_{\Tau,\{i,j,k\}}(u)
    =
    \widetilde{{\cal G}}^{\rm tri.,(2)}_{\Tau,\{i,j,k\}}(u)
    + \int_0^{1} {\rm d}\xi
      \int_0^{2\pi} {\rm d}\phi_{q} \;
      \widetilde{{\cal F}}^{\rm tri.,(2),(R)}_{\Tau,\{i,j,k\}}
      (u,\xi,\phi_{q},t^\pm)\, .
\end{align}
The observable-independent contribution is given by
\begin{align}
    \widetilde{{\cal G}}^{\rm tri.,(2),(R)}_{\Tau,\{i,j,k\}}(u)
    ={}& -\frac{4\pi}{3} \left( L_{ij}^3 +
    2 \pi^2 L_{ij} +\theta(n_i \cdot n_k - n_j \cdot n_k ) \ln^3 \frac{n_i \cdot n_k}{n_j \cdot n_k}
 \right)\,,
\end{align}
with $L_{ij}$ defined in Eq.~(\ref{eqn:Gammas-def}), while the dependence on the observable is encoded in the function
$\widetilde{{\cal F}}^{\rm tri.,(2),(R)}_{\Tau,\{i,j,k\}}
(u,\xi,\phi_{q},t^\pm)$, which can be safely integrated numerically over $\xi$ and $\phi_{q}$. Explicitly, it reads
\begin{align}
&\widetilde{{\cal F}}^{\rm tri.,(2),(R)}_{\Tau,\{i,j,k\}}
(u,\xi,\phi_{q},t^\pm)
= -\frac{4}{\xi}
\bigg[
    L_{jk}
    \ln\!\left(
        t_{N,\{ij\}}^{+}(\xi,\phi_{q})
        t_{N,\{ij\}}^{-}(\xi,\phi_{q})
    \right)
\\
&
    ~~~~~+\ln \left(t_{N,\{ij\}}^{+}(\xi,\phi_{q})\right)
    \ln\!\left(  \frac{\xi\,t_{N,\{ij\}}^{+}(\xi,\phi_{q})}{\,w_k(1,\xi,0,\phi_q)}\right)
    +\ln \left(t_{N,\{ij\}}^{-}(\xi,\phi_{q})\right)
    \ln\!\left( \frac{\xi^2\, t_{N,\{ij\}}^{-}(\xi,\phi_{q})}{\, w_k(\xi,1,0,\phi_q)}\right)
\bigg] .\nn
\end{align}
The representation~\eqref{eq:triple_renorm} of the tripole correction is a new result of this work, and to the best of our knowledge is the most compact one available in the literature.

\subsection{Numerical results for benchmark phase-space points}

In the following, we discuss numerical results for the
$N$-Jettiness soft function obtained using the method presented in this article.
In particular, we perform the comparison with the results in the literature for 0-, 1-, 2-, and 3-Jettiness, and we provide new results for 4- and 5-Jettiness
for selected benchmark points. In particular, we report results for the non-logarithmic terms in Laplace space to facilitate the comparison to the predictions of Ref.~\cite{Bell:2023yso,Agarwal:2024gws}, and therefore all the numerical results presented in this section are obtained by setting $L_{ij}\to 0$, with $L_{ij}$ being defined in Eq.~\eqref{eqn:Gammas-def}.

Following Refs~\cite{Bell:2023yso,Agarwal:2024gws}, we consider the $pp$ case in Eq.~\eqref{eq:2.1}, and choose  the directions of the incoming legs as
\begin{align}
    n_{1,2} = (1,0,0,\pm1) \,.
\end{align}
In this frame, the remaining directions can be parametrized as
\begin{align}
    n_{i}=(1,\sin\theta_{1i}\cos\phi_i,\sin\theta_{1i}\sin\phi_i,\cos\theta_{1i}) \,, ~~ i=3,\dots,N+2\,.
\end{align}
In addition, we choose the $x$-axis in such a way that $\phi_3=0$. Benchmark values for the soft function are obtained for the following phase-space points:
\begin{align}\label{eq:benchmark_pts}
    N=1: {}& ~~
    \theta_{13}=\frac{12\pi}{25}\,,\\
    N=2: {}& ~~
    \theta_{13}=\frac{6\pi}{25}\,,\theta_{14}=\frac{13\pi}{25}\,,
    \phi_4=\frac{\pi}{5}\,,\\
    N=3: {}& ~~
    \theta_{13}=\frac{3\pi}{10}\,,\theta_{14}=\frac{6\pi}{10}\,,\theta_{15} =\frac{9\pi}{10}\,,
    \phi_4=\frac{3\pi}{5}\,,\phi_5=\frac{6\pi}{5}\,,
    \label{eq:benchmark_pts-3jettiness}
    \\
    N=4: {}& ~~
    \theta_{13}=\frac{12 \pi}{25}\,,\theta_{14}=\frac{11 \pi}{25}\,,\theta_{15} =\frac{18 \pi}{25}\,,\theta_{16} =\frac{20 \pi}{25}, \nonumber \\ 
    {}& ~~ \phi_4=\frac{8 \pi}{25}\,,\phi_5=\frac{11 \pi}{25}\,,\phi_6=\frac{22 \pi}{25}\,, \\
    N=5: {}& ~~
    \theta_{13}=\frac{5\pi}{25}\,,\theta_{14}=\frac{10\pi}{25}\,,\theta_{15} =\frac{9\pi}{25}\,,\theta_{16} =\frac{10\pi}{25}\,,\theta_{17} =\frac{13\pi}{25}, \nonumber \\ 
    {}& ~~ \phi_4=\frac{21\pi}{25}\,,\phi_5=\frac{2\pi}{25}\,,\phi_6=\frac{4\pi}{25}\,,\phi_7=\frac{22\pi}{25}\,.
\end{align}
For the case of 0-Jettiness there is only one dipole;  we present the results for this case in Table~\ref{tab:dip-0JN} finding exquisite agreement with the results in the literature \cite{Kelley:2011ng,Hornig:2011iu,Monni:2011gb}.
\begin{table}[ht]
   \centering
    \begin{tabular}{||c|l|l|l||l|l|l||}
     \hline \hline
    \multirow{2}{*}{} &\multicolumn{3}{c||}{Gluons ($C_A$ term)} &\multicolumn{3}{c||}{Quarks ($T_F n_f$ term)} \\
    \cline{2-7}
    & \multicolumn{1}{c|}{$\tilde{s}^{(2)}_{\TauZero}$} &
      \multicolumn{1}{c|}{$\tilde{s}^{(2)}_{\TauZeroBar}$} &
      \multicolumn{1}{c||}{$\Delta \tilde{s}^{(2)}_{\TauZero}$} &
      \multicolumn{1}{c|}{$\tilde{s}^{(2)}_{\TauZero}$} &
      \multicolumn{1}{c|}{$\tilde{s}^{(2)}_{\TauZeroBar}$} &
      \multicolumn{1}{c||}{$\Delta \tilde{s}^{(2)}_{\TauZero}$}  \\
    \hline \hline
Numerical & $28.254(7)$ & $34.70461$ & $-6.451(7)$ & $-21.6954(3)$ & $-19.06462$ & $-2.6308(3)$ \\
    \hline
Analytic & $28.2495$ & $34.70461$ & $-6.4551$ & $-21.6953$ & $-19.06462$ & $-2.63063$ \\
    \hline
    \end{tabular}
    \caption{Comparison for the $0$-Jettiness soft function constant in Laplace space obtained with the method of this paper and the analytic determination of Refs~\cite{Kelley:2011ng,Hornig:2011iu,Monni:2011gb}. The result for inclusive soft function, $\tilde{s}^{(2)}_{\TauZeroBar}$, is obtained by taking the Laplace transform of the expression calculated analytically in \sec{0Jettiness}.}
\label{tab:dip-0JN}
\end{table}

\begin{figure}
\includegraphics[width=\textwidth]{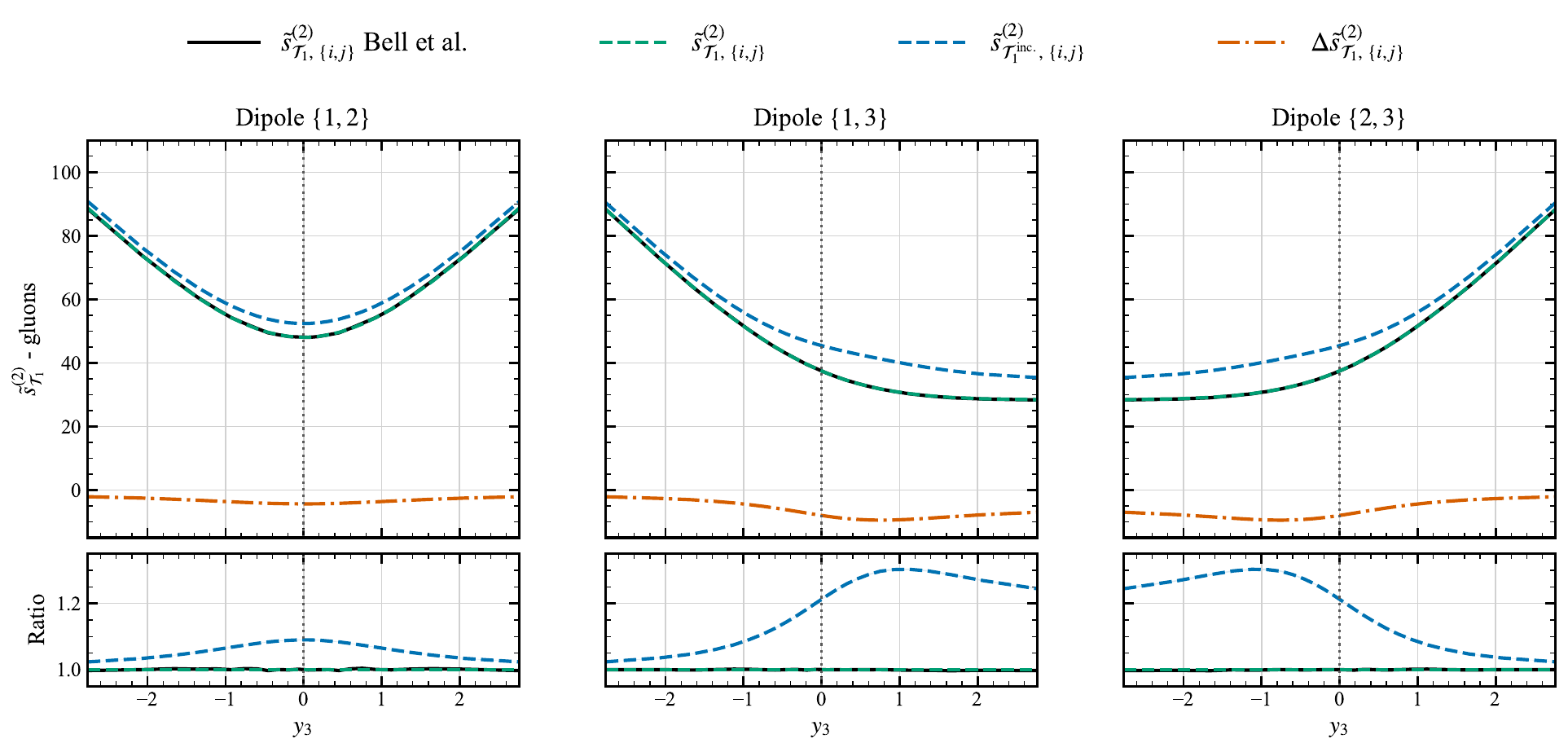}
\caption{\label{fig:1JN-CA} Comparison of the results for the gluonic final state contribution to the dipole correction to the $1$-Jettiness soft function for different values of the jet rapidity $y_3=1/2\ln \left[(1+\cos\theta_{13})/(1-\cos\theta_{13})\right]$ and for the three dipoles $\{1,2\}$, $\{1,3\}$, and $\{2,3\}$. The plot compares the results obtained with the method presented in this article (green dashed line) with those obtained in Ref.~\cite{Bell:2023yso} (black line). The separation in the inclusive, $\tilde{s}^{(2)}_{\TauNinc{1},\,\{i,j\}}$ (blue dashed line) and $\Delta \tilde{s}^{(2)}_{\TauN{1},\,\{i,j\}}$ (orange, dot-dashed line) contributions is also shown. The ratios of the various curves to the full results obtained with our method are shown in the bottom panel.}
\end{figure}

\begin{figure}
\includegraphics[width=\textwidth]{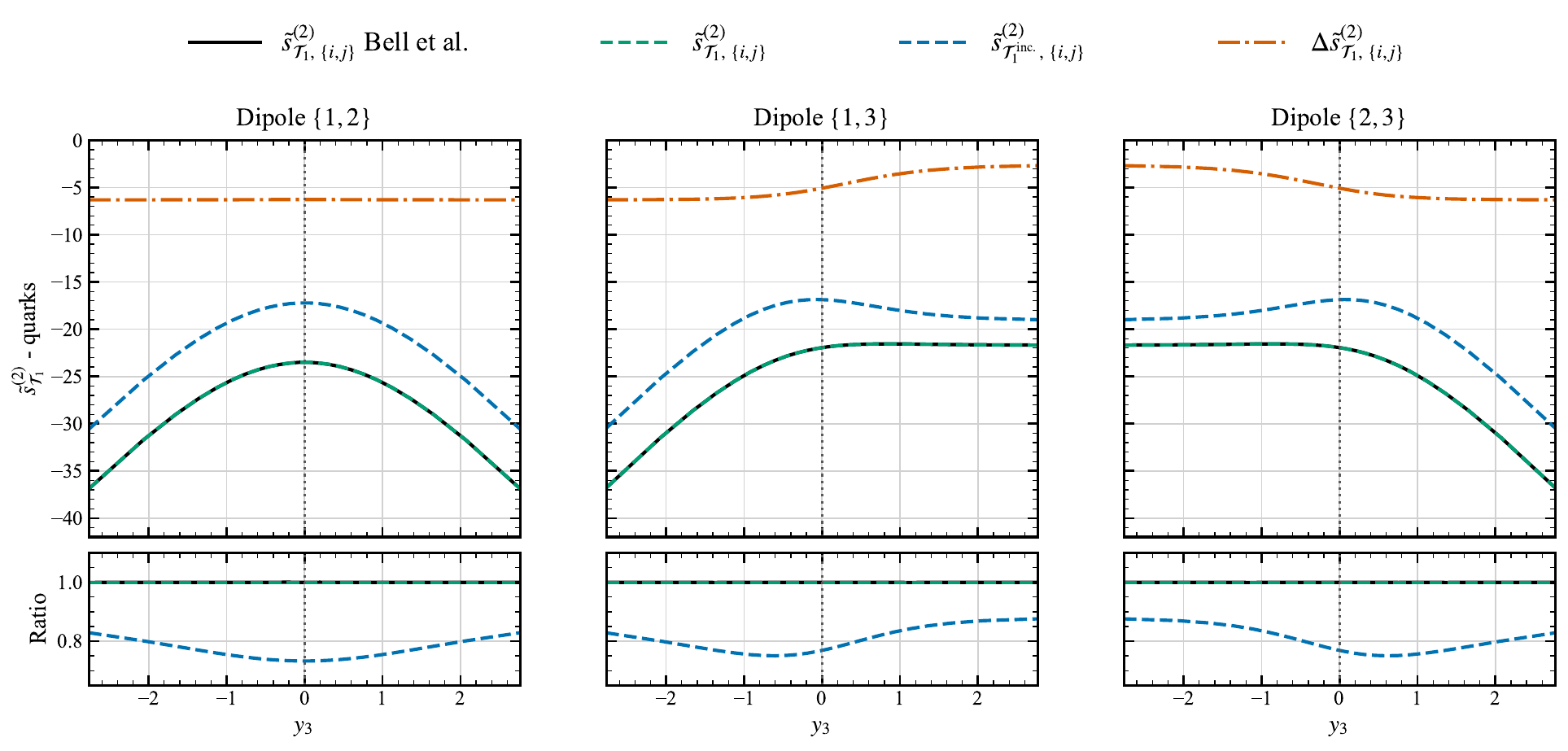}
\caption{\label{fig:1JN-nf} Same as Fig.~\ref{fig:1JN-CA} for the fermionic contributions.}
\end{figure}

We then move to the evaluation of the $1$-Jettiness soft function. In this case, the result is a function  of a single angular variable, that we choose to be the pseudo-rapidity of the jet $y_3=1/2\ln \left[(1+\cos\theta_{13})/(1-\cos\theta_{13})\right]$.
The corresponding results of the numerical calculation are displayed in Figs~\ref{fig:1JN-CA} and~\ref{fig:1JN-nf} for gluonic and fermionic final-state channels, respectively. For comparison, we also report the results of Ref.~\cite{Bell:2023yso}, with which we find excellent agreement.
In addition, the plots separately  show  the contribution from the inclusive approximation to the $1$-Jettiness soft function, computed in Section~\ref{sec:NJ-inclusive}, and the non-inclusive correction $\Delta \tilde{s}^{(2)}_{\Tau,\,\{i,j\}}$ (cf.~Section~\ref{sec:NJettiness-num-deltaij}).
We observe that the inclusive contribution captures most of the radiative corrections to the full result.

Tables~\ref{tab:dip-2JN},~\ref{tab:dip-3JN} report the numerical results for $2$- and $3$-Jettiness at the benchmark points defined in  Eq.~\eqref{eq:benchmark_pts}, 
which are also used in Refs~\cite{Agarwal:2024gws,Bell:2023yso}.
In addition, we also report new results for $4$- and $5$-Jettiness at the benchmark points given in Eq.~\eqref{eq:benchmark_pts}.
The latter results are also reported in Tables~\ref{tab:dip-4JN},~\ref{tab:dip-5JN} for comparison. The numerical evaluation of the full soft function is rather fast.
For reference, on a single thread (MacBook Pro M3), the average evaluation time per dipole (for the combination of the gluonic and fermionic final-state channels) needed to achieve a precision of $1\%$ and $0.1\%$ for 1-,2-,3-,4-,5-Jettiness is ${\cal O}(0.1s)$-${\cal O}(1s)$ and ${\cal O}(1s)$-${\cal O}(10s)$, respectively.
The increase in the evaluation time with the number of jets is due to the growth in the number of invariants that enter the definition of the observable in the minimum of Eq.~\eqref{eq:observable}.

Our numerical implementation relies on a relatively simple parametrization
  of the two-emission phase space, combined with the standard adaptive Monte
  Carlo algorithm Vegas~\cite{Lepage:1977sw}, as implemented in the Cuba
  library~\cite{Hahn:2004fe}. Consequently, we expect that significant
  opportunities remain for further optimizations and performance improvements.
  We leave a systematic investigation of these aspects to future work, with the
  goal of developing a robust, user-friendly public code aimed at applications
  involving slicing methods and higher-order matching. Given that the
  computational timings reported here are already comparable to those of complex
  two-loop virtual amplitudes evaluated during phase-space integration, we
  anticipate that the soft function can likewise be computed on the fly. This
  would eliminate the need for multidimensional interpolation grids and the
  associated technical complications.
\begin{table}[ht]
   \centering
    \begin{tabular}{||c|r|r|r||r|r|r||}
     \hline \hline
    \multirow{2}{*}{Dip.} &\multicolumn{3}{c||}{Gluons} &\multicolumn{3}{c||}{Quarks} \\
    \cline{2-7}
    & \multicolumn{1}{c|}{$\tilde{s}^{(2)}_{\TauN{1},\,\{i,j\}}$} &
      \multicolumn{1}{c|}{Ref.~\cite{Agarwal:2024gws}} &
      \multicolumn{1}{c||}{Ref.~\cite{Bell:2023yso}} & \multicolumn{1}{c|}{$\tilde{s}^{(2)}_{\TauN{1},\,\{i,j\}}$} &
      \multicolumn{1}{c|}{Ref.~\cite{Agarwal:2024gws}} &
      \multicolumn{1}{c||}{Ref.~\cite{Bell:2023yso}}  \\
    \hline \hline
     12 & 48.10(4) & 48.10(1)  & 48.04(7)  &
          -23.504(3) & -23.504(1) &-23.503(10)  \\
    \cline{1-7}
     13 & 36.84(5) & 36.81(2) & 36.82(4)  &
          -21.874(3)  & -21.871(3)  & -21.875(9)  \\
    \cline{1-7}
     23 & 38.15(5) & 38.14(1) & 38.13(5)  &
           -22.032(3)   & -22.039(2)   & -22.031(9) \\
    \hline
    \end{tabular}
    \caption{Numerical results for the dipole contribution to the $1$-Jettiness soft function for the benchmark kinematic point in Eq.~\eqref{eq:benchmark_pts} compared with those reported in Refs~\cite{Agarwal:2024gws,Bell:2023yso}.
    }
\label{tab:dip-1JN}
\end{table}

\begin{table}[ht]
   \centering
    \begin{tabular}{||c|r|r|r||r|r|r||}
     \hline \hline
    \multirow{2}{*}{Dip.} &\multicolumn{3}{c||}{Gluons} &\multicolumn{3}{c||}{Quarks} \\
    \cline{2-7}
    & \multicolumn{1}{c|}{$\tilde{s}^{(2)}_{\TauN{2},\,\{i,j\}}$} &
      \multicolumn{1}{c|}{Ref.~\cite{Agarwal:2024gws}} &
      \multicolumn{1}{c||}{Ref.~\cite{Bell:2023yso}} & \multicolumn{1}{c|}{$\tilde{s}^{(2)}_{\TauN{2},\,\{i,j\}}$} &
      \multicolumn{1}{c|}{Ref.~\cite{Agarwal:2024gws}} &
      \multicolumn{1}{c||}{Ref.~\cite{Bell:2023yso}}  \\
    \hline \hline
     12 & 71.16(5)  & 71.15(5) & 71.11(12) &
           -27.840(6) & -27.837(1) & -27.841(11)  \\
    \cline{1-7}
     13 & 36.25(5) & 36.27(2) & 36.17(7) &
          -21.722(4) & -21.719(5) & -21.724(9) \\
    \hline
     23 & 75.80(6) & 75.79(1)  & 75.62(9) &
          -27.809(7) & -27.804(2)  & -27.807(11) \\
    \hline
     14 & 65.51(6) & 65.48(2) &  65.38(9) &
          -25.656(6) &-25.660(3)  & -25.666(10)\\
    \hline
     24 & 46.25(5) & 46.25(1)  & 46.15(6)  &
          -22.937(4) &-22.933(5) & -22.908(9) \\
    \hline
     34 & 44.81(6) & 44.86(2) & 44.72(9)  &
          -22.517(5) &-22.513(4)  & -22.518(9) \\
    \hline
    \end{tabular}
    \caption{Same as in Table~\ref{tab:dip-1JN} for the $2$-Jettiness soft function.
    }
\label{tab:dip-2JN}
\end{table}

\begin{table}[ht]
   \centering
    \begin{tabular}{||c|r|r|r||r|r|r||}
     \hline \hline
    \multirow{2}{*}{Dip.} &\multicolumn{3}{c||}{Gluons} &\multicolumn{3}{c||}{Quarks} \\
    \cline{2-7}
    & \multicolumn{1}{c|}{$\tilde{s}^{(2)}_{\TauN{3},\,\{i,j\}}$} &
      \multicolumn{1}{c|}{Ref.~\cite{Agarwal:2024gws}} &
      \multicolumn{1}{c||}{Ref.~\cite{Bell:2023yso}} & \multicolumn{1}{c|}{$\tilde{s}^{(2)}_{\TauN{3},\,\{i,j\}}$} &
      \multicolumn{1}{c|}{Ref.~\cite{Agarwal:2024gws}} &
      \multicolumn{1}{c||}{Ref.~\cite{Bell:2023yso}}  \\
    \hline \hline
     12 & 116.17(7) & 116.20(1) & 116.20(16)  &
          -36.257(14) & -36.249(1) & -36.244(9)  \\
    \cline{1-7}
     13 & 37.65(5) & 38.13(3) & 37.63(3) &
          -21.728(4) & -21.717(7) & -21.732(5)\\
    \hline
     14 & 63.60(6) & 63.63(1) & 63.66(6) &
           -25.195(6)  &-25.189(3)  & -25.192(6) \\
    \hline
     15 & 107.14(7) & 107.17(1) & 106.99(12) &
           -35.271(12)  & -35.268(1)  & -35.256(9)\\
    \hline
     23 & 97.09(7) & 97.11(1)  & 96.97(10)  &
          -32.877(12)  & -32.875(2)  & -32.872(8)\\
    \hline
     24 & 67.37(7) & 67.36(2)  & 67.51(8)  &
          -26.817(10) & -26.821(3) & -26.815(7)\\
    \hline
     25 & 30.83(5) & 30.87(3)  & 30.73(4)  &
           -21.555(3) & -21.561(9) & -21.561(5)\\
    \hline
     34 & 69.45(6) & 69.43(1) & 69.24(7)   &
          -25.852(7)  & -25.854(2)  & -25.861(6) \\
    \hline
     35 & 106.12(7) & 106.13(2) & 105.97(13)   &
         -34.794(13)  & -34.799(2)  & -34.796(8)\\
    \hline
     45 & 74.43(7) & 74.45(2) & 74.36(9)  &
       -28.244(10)  &  -28.247(4)  & -28.251(7)\\
    \hline
    \end{tabular}
    \caption{Same as Table~\ref{tab:dip-1JN} for the $3$-Jettiness soft function.
    }
\label{tab:dip-3JN}
\end{table}

\begin{table}[ht]
   \centering
    \begin{tabular}{||c|r|r|r||r|r|r||}
     \hline \hline
    \multirow{2}{*}{Dip.} &\multicolumn{3}{c||}{Gluons} &\multicolumn{3}{c||}{Quarks} \\
    \cline{2-7}
    & \multicolumn{1}{c|}{$\tilde{s}^{(2)}_{\TauN{4},\,\{i,j\}}$} &
      \multicolumn{1}{c|}{$\tilde{s}^{(2)}_{\TauNinc{4},\,\{i,j\}}$} &
      \multicolumn{1}{c||}{$\Delta \tilde{s}^{(2)}_{\TauN{4},\,\{i,j\}}$} & \multicolumn{1}{c|}{$\tilde{s}^{(2)}_{\TauN{4},\,\{i,j\}}$} &
      \multicolumn{1}{c|}{$\tilde{s}^{(2)}_{\TauNinc{4},\,\{i,j\}}$} &
      \multicolumn{1}{c||}{$\Delta \tilde{s}^{(2)}_{\TauN{4},\,\{i,j\}}$}  \\
    \hline \hline
12 & $116.74(7)$ & $117.79(4)$ & $-1.04(6)$ & $-36.03(1)$ & $-19.706(8)$ & $-16.33(1)$ \\
    \hline
13 & $62.58(6)$ & $74.35(2)$ & $-11.76(6)$ & $-24.875(6)$ & $-14.791(2)$ & $-10.084(6)$ \\
\hline
14 & $62.04(7)$ & $74.47(2)$ & $-12.43(6)$ & $-24.671(6)$ & $-14.414(3)$ & $-10.257(6)$ \\
\hline
15 & $109.14(7)$ & $108.50(3)$ & $0.63(6)$ & $-33.27(1)$ & $-17.023(7)$ & $-16.244(9)$ \\
\hline
16 & $98.69(7)$ & $107.96(3)$ & $-9.28(6)$ & $-32.99(1)$ & $-18.801(5)$ & $-14.193(9)$ \\
\hline
23 & $84.55(7)$ & $95.68(3)$ & $-11.14(6)$ & $-29.158(10)$ & $-15.502(5)$ & $-13.656(8)$ \\
\hline
24 & $110.55(8)$ & $109.32(4)$ & $1.24(8)$ & $-32.87(1)$ & $-15.848(7)$ & $-17.02(1)$ \\
\hline
25 & $56.69(7)$ & $70.52(2)$ & $-13.83(7)$ & $-23.428(7)$ & $-13.51(3)$ & $-9.919(6)$ \\
\hline
26 & $40.92(6)$ & $56.055(9)$ & $-15.14(6)$ & $-21.874(4)$ & $-15.449(1)$ & $-6.425(4)$ \\
\hline
34 & $52.00(7)$ & $66.90(2)$ & $-14.9(7)$ & $-22.852(6)$ & $-13.769(2)$ & $-9.083(5)$ \\
\hline
35 & $83.66(8)$ & $90.11(3)$ & $-6.45(7)$ & $-27.733(10)$ & $-13.592(5)$ & $-14.141(8)$ \\
\hline
36 & $118.22(8)$ & $117.07(3)$ & $1.14(7)$ & $-35.49(1)$ & $-18.511(7)$ & $-16.98(1)$ \\
\hline
45 & $54.52(7)$ & $71.44(2)$ & $-16.93(6)$ & $-23.312(6)$ & $-13.641(2)$ & $-9.67(5)$ \\
\hline
46 & $108.49(8)$ & $110.59(4)$ & $-2.1(7)$ & $-33.16(1)$ & $-16.698(7)$ & $-16.47(1)$ \\
\hline
56 & $58.37(7)$ & $71.84(2)$ & $-13.47(7)$ & $-23.837(7)$ & $-13.849(3)$ & $-9.988(6)$ \\
    \hline
    \end{tabular}
    \caption{Numerical results for the gluonic and fermionic dipole contributions to $4$-Jettiness soft function for the benchmark kinematic point in Eq.~\eqref{eq:benchmark_pts}. The decomposition into the inclusive, $\tilde{s}^{(2)}_{\TauNinc{4},\,\{i,j\}}$, and delta, $\Delta \tilde{s}^{(2)}_{\TauN{4},\,\{i,j\}}$, contributions is also reported.
    }
\label{tab:dip-4JN}
\end{table}

\begin{table}[ht]
   \centering
    \begin{tabular}{||c|r|r|r||r|r|r||}
     \hline \hline
    \multirow{2}{*}{Dip.} &\multicolumn{3}{c||}{Gluons} &\multicolumn{3}{c||}{Quarks} \\
    \cline{2-7}
    & \multicolumn{1}{c|}{$\tilde{s}^{(2)}_{\TauN{5},\,\{i,j\}}$} &
      \multicolumn{1}{c|}{$\tilde{s}^{(2)}_{\TauNinc{5},\,\{i,j\}}$} &
      \multicolumn{1}{c||}{$\Delta \tilde{s}^{(2)}_{\TauN{5},\,\{i,j\}}$} & \multicolumn{1}{c|}{$\tilde{s}^{(2)}_{\TauN{5},\,\{i,j\}}$} &
      \multicolumn{1}{c|}{$\tilde{s}^{(2)}_{\TauNinc{5},\,\{i,j\}}$} &
      \multicolumn{1}{c||}{$\Delta \tilde{s}^{(2)}_{\TauN{5},\,\{i,j\}}$}  \\
    \hline \hline
12 & $117.86(7)$ & $118.63(3)$ & $-0.77(6)$ & $-34.91(1)$ & $-19.737(7)$ & $-15.176(9)$ \\
\hline
13 & $46.48(6)$ & $63.14(1)$ & $-16.66(6)$ & $-22.435(5)$ & $-15.548(2)$ & $-6.887(5)$ \\
\hline
14 & $78.12(7)$ & $92.40(3)$ & $-14.27(7)$ & $-27.723(10)$ & $-16.382(4)$ & $-11.341(8)$ \\
\hline
15 & $106.14(9)$ & $111.01(4)$ & $-4.87(8)$ & $-32.48(1)$ & $-18.295(8)$ & $-14.18(1)$ \\
\hline
16 & $109.55(9)$ & $117.34(3)$ & $-7.79(8)$ & $-34.1(1)$ & $-19.80(7)$ & $-14.3(1)$ \\
\hline
17 & $107.03(8)$ & $113.42(3)$ & $-6.39(7)$ & $-33.0(1)$ & $-18.967(6)$ & $-14.032(9)$ \\
\hline
23 & $127.28(7)$ & $127.99(4)$ & $-0.71(6)$ & $-37.56(1)$ & $-21.534(7)$ & $-16.022(10)$ \\
\hline
24 & $99.26(8)$ & $103.74(3)$ & $-4.48(7)$ & $-31.66(1)$ & $-18.997(7)$ & $-12.659(8)$ \\
\hline
25 & $122.88(8)$ & $128.24(4)$ & $-5.36(7)$ & $-38.09(2)$ & $-23.17(9)$ & $-14.92(1)$ \\
\hline
26 & $104.94(8)$ & $115.32(4)$ & $-10.38(7)$ & $-34.38(1)$ & $-21.149(9)$ & $-13.23(1)$ \\
\hline
27 & $71.42(6)$ & $84.26(2)$ & $-12.84(6)$ & $-27.048(8)$ & $-17.315(3)$ & $-9.734(7)$ \\
\hline
34 & $132.85(8)$ & $132.37(4)$ & $0.48(7)$ & $-37.78(1)$ & $-20.378(8)$ & $-17.4(1)$ \\
\hline
35 & $53.24(6)$ & $68.56(2)$ & $-15.32(6)$ & $-23.609(7)$ & $-15.556(3)$ & $-8.053(7)$ \\
\hline
36 & $77.16(7)$ & $88.06(3)$ & $-10.91(7)$ & $-27.395(9)$ & $-16.708(5)$ & $-10.687(8)$ \\
\hline
37 & $153.80(9)$ & $149.22(5)$ & $4.59(8)$ & $-42.89(2)$ & $-24.034(9)$ & $-18.85(1)$ \\
\hline
45 & $166.83(9)$ & $160.94(5)$ & $5.89(8)$ & $-46.36(2)$ & $-26.504(10)$ & $-19.86(2)$ \\
\hline
46 & $142.16(9)$ & $145.81(4)$ & $-3.64(8)$ & $-41.66(2)$ & $-23.882(9)$ & $-17.78(2)$ \\
\hline
47 & $33.36(5)$ & $46.776(4)$ & $-13.41(5)$ & $-21.538(3)$ & $-17.4058(5)$ & $-4.132(3)$ \\
\hline
56 & $33.83(5)$ & $47.418(6)$ & $-13.59(5)$ & $-21.565(4)$ & $-17.2284(8)$ & $-4.336(4)$ \\
\hline
57 & $178.31(10)$ & $170.85(5)$ & $7.46(8)$ & $-49.52(2)$ & $-29.04(1)$ & $-20.47(2)$ \\
\hline
67 & $153.52(9)$ & $154.36(5)$ & $-0.84(8)$ & $-44.39(2)$ & $-25.91(1)$ & $-18.47(2)$ \\
    \hline
    \end{tabular}
    \caption{Same as Table~\ref{tab:dip-4JN} for the $5$-Jettiness soft function.
    }
\label{tab:dip-5JN}
\end{table}

Finally, we validated the new representation of the tripole contribution presented in Eq.~\eqref{eq:triple_renorm}. In Table~\ref{tab:triple-correlators-num}, we compare our predictions for the tripole correction to the $3$-Jettiness NNLO soft function to those of Refs~\cite{Bell:2023yso,Agarwal:2024gws}, finding very good agreement.

\begin{table}[ht]
   \centering
    \begin{tabular}{||c|l|l|l|l||}
     \hline \hline
     & \multicolumn{1}{c|}{$\tilde{c}_{\text{tripoles}}^{(2,124)}$} & \multicolumn{1}{c|}{$\tilde{c}_{\text{tripoles}}^{(2,125)}$} & \multicolumn{1}{c|}{$\tilde{c}_{\text{tripoles}}^{(2,145)}$} & \multicolumn{1}{c||}{$\tilde{c}_{\text{tripoles}}^{(2,245)}$} \\
    \hline \hline
     $\tilde{c}_{\text{tripoles}}$ & -683.23 $\pm$ 0.01 & -2203.50 $\pm$ 0.01  & -6.312 $\pm$ 0.009 & -0.840 $\pm$ 0.008  \\
     \hline \hline
     Ref.~\cite{Agarwal:2024gws} & -683.25 $\pm$ 0.01 & -2203.3 $\pm$ 0.2  & -6.324 $\pm$ 0.004 & -0.837 $\pm$ 0.008  \\
    \cline{1-5}
     Ref.~\cite{Bell:2023yso} & -683.23 $\pm$ 0.04 & -2203.5 $\pm$ 0.1 & -6.325 $\pm$ 0.04 & -0.830 $\pm$ 0.039\\
    \hline
    \end{tabular}
    \caption{Results for the non-logarithmically enhanced tripole contributions $\tilde{c}_{\text{tripoles}}^{(2,ijk)}$ to the $3$-Jettiness soft function at the benchmark point specified in Eq.~\eqref{eq:benchmark_pts-3jettiness}, compared to results from the literature. Following Ref.~\cite{Bell:2023yso} (see Eqs (2.64,3.13)), we define $\tilde{c}_{\text{tripoles}}^{(2,ijk)}$ as the coefficients of the tripole contribution in a color basis where we replace $\colT{3}=-(\colT{1}+\colT{2}+\colT{4}+\colT{5})$ using color conservation, such that there are only four independent color structures ${\boldsymbol F}_{ijk} $ with $(ijk)\in\{(124),(125),(145),(245)\}$.}
    \label{tab:triple-correlators-num}
\end{table}

\section{Outlook}
\label{sec:Conclusions}
%
In this paper, we have presented a novel strategy to simplify the calculation of $N$-Jettiness soft functions at high orders in perturbation theory.
By isolating the contribution of the dipole correlator for an arbitrary number of hard emitters, we expressed this contribution to the soft function as the sum of an analytically computable piece and a finite remainder that can be evaluated numerically with greatly reduced complexity.
At NNLO, the remainder requires no subtraction, while at N$^3$LO we expect its structure to be analogous to an NLO cross-section computation, which would necessitate only simple NLO-type subtractions~\cite{Buonocore:2026FutureJettiness}.
Looking ahead, a similar approach can be envisioned to tackle the contribution from higher-point color correlators to the $N$-Jettiness soft function, that we will pursue in future work.

We have applied our method to the $N$-Jettiness soft function and provided numerical results at NNLO for $N\leq 5$, demonstrating both the flexibility and efficiency of our approach.
Our framework provides a scalable pathway for extending $N$-Jettiness subtraction methods to processes with jets in the final state at N$^3$LO, enabling us to tackle one of the key bottlenecks in precision QCD calculations for the LHC and future colliders.

\section*{Acknowledgments}
KM is  partially supported by the Deutsche Forschungsgemeinschaft (DFG, German Research Foundation) under grant 396021762 - TRR 257.
The work of AP is supported by the ERC Consolidator Grant 101169614.
The work of LB, MD, PM and GV is funded by the European Union (ERC, grant agreement No. 101044599). Views and opinions expressed are however those of the authors only and do not necessarily reflect those of the European Union or the European Research Council Executive Agency. Neither the European Union nor the granting authority can be held responsible for them.
\appendix

\section{Eikonal functions}
\label{app:N2LO-eikonals}

For completeness, we provide NNLO eikonal functions that we use to compute the relevant contributions to the
$N$-Jettiness soft function.

\subsection{Single-real contribution}
\begin{align}
    \label{eq:single-soft-eikonal}
    S_{ij}(k_1) = \frac{n_i \cdot n_j}{(n_i\cdot k_1)(n_j\cdot k_1)}\,.
\end{align}

\subsection{Real-virtual contribution}
The color-stripped real-virtual tripole contribution to soft matrix element reads~\cite{Catani:2000pi}
\begin{align}
\label{eq:Ms-RV-tripole}
 \mathcal{M}^{(1)}_{s,\{ \{i,j\},k\}}(k_1) = S_{ij}(k_1) (S_{jk}(k_1))^{\epsilon}\,.
\end{align}

\subsection{Double-real contribution}
We decompose the maximally non-Abelian part of the double-real tree-level soft matrix element into a gluon- and a quark-emission piece
\begin{align}
    \label{eq:Ms-RR-nab}
    \mathcal{M}^{(0)}_{s,\{i,j\}}(k_1,k_2) = - \frac1{2} C_A \mathcal{M}^{(0),gg}_{s,\{i,j\}}(k_1,k_2) + T_R n_f \mathcal{M}^{(0),q\bar{q}}_{s,\{i,j\}}(k_1,k_2) \,,
\end{align}
where we included the $1/2$ symmetry factor for identical gluons in the final state and defined 
\begin{align}
 \mathcal{M}^{(0),ff}_{s,\{i,j\}}(k_1,k_2) = 2 S_{ij}^{ff}(k_1,k_2)  - S_{ii}^{ff}(k_1,k_2) - S_{jj}^{ff}(k_1,k_2) \,,\qquad ff=gg,q\bar{q}\,.
\end{align}
The functions $S_{ij}^{gg/q\bar{q}}(k_1,k_2)$ read~\cite{Catani:1999ss}
\be
\begin{split}
S_{ij}^{gg}(k_1,k_2)
&= S_{ij}^{gg, \rm so}(k_1,k_2)
- \frac{2\,n_i \cdot n_j}{
k_1 \cdot k_2 [ n_i \cdot (k_1 + k_2)][n_j \cdot (k_1 + k_2) ]
}
\\
&+ \frac{
(n_i \cdot k_1)(n_j \cdot k_2)+
(n_i \cdot k_2)(n_j \cdot k_1)
}{[ n_i \cdot (k_1 + k_2) n_j \cdot (k_1 + k_2) ]}
\left [
\frac{1-\ep}{(k_1 \cdot k_2)^2}
- \frac{1}{2} S_{ij}^{gg, \rm so}(k_1,k_2)
\right ],
\end{split}
\ee
where
\be
\begin{split}
S_{ij}^{gg, \rm so}(k_1,k_2)
& = \frac{n_i \cdot n_j}{k_1 \cdot k_2}
\left (
\frac{1}{(n_i \cdot k_1)(n_j \cdot k_2)}
+
\frac{1}{(n_i \cdot k_2)(n_j \cdot k_1)}
\right )
\\
& -\frac{(n_i \cdot n_j)^2}{(n_i \cdot k_1)(n_j \cdot k_1) (n_i \cdot k_2) (n_j \cdot k_2) } \,,
\end{split}
\ee
and
\be
\begin{split}
S_{ij}^{q\bar q}(k_1,k_2)
=
\frac{(n_i \cdot k_1)(n_j \cdot k_2)+
(n_i \cdot k_2)(n_j \cdot k_1)
- (n_i \cdot n_j) (k_1 \cdot k_2)
}{(k_1 \cdot k_2)^2
[ n_i \cdot (k_1 + k_2)][n_j \cdot (k_1 + k_2) ]
}.
\end{split}
\ee

\section{Renormalization of the soft function up to NNLO}
\label{app:renorm}
In this appendix, we describe the renormalization of UV divergences in the $\Tau$ soft function through NNLO.
With reference to Eq.~\eqref{eq:renorm}, the renormalized soft function $\widetilde {\boldsymbol  {\cal S}}_{\Tau}^{(R)}$ in Laplace space is related to its bare counterpart $\widetilde {\boldsymbol  {\cal S}}_{\Tau}$ by the relation
\begin{equation}
\widetilde {\boldsymbol  {\cal S}}_{\Tau} (u) =    {\boldsymbol Z} \widetilde {\boldsymbol  {\cal S}}^{(R)}_{\Tau} (u) {\boldsymbol Z} ^\dagger\,.
\end{equation}
where ${\boldsymbol Z}$ is a matrix in color space.
It is convenient to expand Eq.~\eqref{eq:renorm} to the target perturbative order to calculate directly renormalized quantities whenever possible. We then use the perturbative expansions in Eqs~\eqref{eq:expansions} to recast the coefficients of the renormalized soft function up to NNLO as follows
\begin{align}\label{eq:renorm-expanded}
\widetilde {\boldsymbol  {\cal S}}^{(R),\,(1)}_{\Tau} &= \widetilde {\boldsymbol  {\cal S}}^{(1)}_{\Tau} - {\boldsymbol Z}^{(1)} - {\boldsymbol Z}^{\dagger,\,(1)},\\
\widetilde {\boldsymbol  {\cal S}}^{(R),\,(2)}_{\Tau} &= \widetilde {\boldsymbol  {\cal S}}^{(2)}_{\Tau} - {\boldsymbol Z}^{(2)} - {\boldsymbol Z}^{\dagger,\,(2)} + {\boldsymbol Z}^{(1)}{\boldsymbol Z}^{(1)} + {\boldsymbol Z}^{\dagger,\,(1)} {\boldsymbol Z}^{\dagger,\,(1)} \notag \\ & ~~~~~~~~~~- {\boldsymbol Z}^{(1)} \widetilde {\boldsymbol  {\cal S}}^{(1)}_{\Tau} - \widetilde {\boldsymbol  {\cal S}}^{(1)}_{\Tau} {\boldsymbol Z}^{\dagger,\,(1)} + {\boldsymbol Z}^{(1)} {\boldsymbol Z}^{\dagger,\,(1)}.\notag
\end{align}
It is also convenient to make use of the exponential structure in Eq.~\eqref{eq:softExp}, to rewrite the two-loop bare soft function as
\begin{align}
    \widetilde {\boldsymbol  {\cal S}}^{(2)}_{\Tau} = \frac{1}{2} \tilde{\boldsymbol s}^{(1)}_{\Tau} \tilde{\boldsymbol s}^{(1)}_{\Tau} + \tilde{\boldsymbol s}^{(2)}_{\Tau}\,.
\end{align}
We recast the renormalization matrix in a form of the following exponential of matrices in color space
\begin{equation}
    {\boldsymbol Z} = \exp\left\{\sum_{n=1}^{\infty}\left(\frac{\alpha_s}{4\pi}\right)^n {\boldsymbol z}^{(n)} \right\}\,,
\end{equation}
so that
\begin{align}
  {\boldsymbol Z}^{(2)} = \frac{1}{2} {\boldsymbol z}^{(1)} {\boldsymbol z}^{(1)} + {\boldsymbol z}^{(2)}\,.
\end{align}
Therefore,
\begin{align}
    \widetilde {\boldsymbol  {\cal S}}^{(R),\,(2)}_{\Tau} = \frac{1}{2} \tilde{\boldsymbol s}^{(R),\,(1)}_{\Tau} \tilde{\boldsymbol s}^{(R),\,(1)}_{\Tau} + \tilde{\boldsymbol s}^{(R),\,(2)}_{\Tau}\,,
\end{align}
with
\begin{align}
    \label{eqn:s2ren-s1bare}
    \tilde{\boldsymbol s}^{(R),\,(2)}_{\Tau}
    ={} & \tilde{\boldsymbol s}^{(2)}_{\Tau} - {\boldsymbol z}^{(2)}-{\boldsymbol z}^{\dagger,\,(2)}+\frac{1}{2}\left[{\boldsymbol z}^{(1)},{\boldsymbol z}^{\dagger,\,(1)}\right]+\frac{1}{2}\left[\tilde{\boldsymbol s}^{(1)}_{\Tau},{\boldsymbol z}^{(1)}-{\boldsymbol z}^{\dagger,\,(1)}\right] \,.
\end{align}

\section{Renormalization of the tripole contribution}
\label{sec:NJettiness-tripole-renorm}

%
Starting from Eq.~\eqref{eqn:s2ren-s1bare}, we work out the renormalization of the tripole correction and show the cancellation of the UV poles.
We start by observing that the two loop contribution to the renormalization constant ${\boldsymbol z}^{(2)}$ is purely dipole-like~\cite{Catani:1998bh,Gardi:2009qi,Becher:2009cu} and, therefore, the poles of the tripole correction are cancelled entirely by the commutators in Eq.~\eqref{eqn:s2ren-s1bare}. To proceed, we write the renormalized NNLO tripole contribution to the soft function as
\begin{align}\label{eq:triple_renorm-0}
    \tilde{\boldsymbol s}^{\rm tri.,(2),(R)}_{\Tau} 
    &= \tilde{\boldsymbol s}^{\rm tri.,(2)}_{\Tau} 
    +\frac{1}{2}\left[{\boldsymbol z}^{(1)},{\boldsymbol z}^{\dagger,\,(1)}\right]
    +\frac{1}{2}\left[{\tilde{\boldsymbol s}}^{(1)}_{\Tau},{\boldsymbol z}^{(1)}-{\boldsymbol z}^{\dagger,\,(1)}\right] \; ,
\end{align}
where
\begin{align}
    {\boldsymbol z}^{(1)} = \sum_{(ij)} \colT{i}\cdot \colT{j} \, z^{(1)}_{\{i,j\}} \,, ~~~ {\boldsymbol z}^{\dagger,(1)} = \sum_{(ij)} \colT{i}\cdot \colT{j} \, z^{*,(1)}_{\{i,j\}}\,,~~~ z^{(1)}_{\{i,j\}} ={} & \left( \frac1{\ep^2} + \frac{L_{ij} + i \pi \lambda_{ij} }{\ep} \right) \,.
\end{align}
To express the commutators in terms of the tripole color structure shown in Eq.~\eqref{eqn:def-Flmn}, we follow steps similar to what was done in Refs~\cite{Devoto:2023rpv,Agarwal:2024gws}. First, we note that after some relabelling of indices the sum over commutators of color dipoles can be written as
\begin{align}
    \label{eqn:comTTTTasHtens}
    \sum_{(il)} \sum_{(jk)} \bigg[ \colT{i}\cdot\colT{l} , \colT{j}\cdot \colT{k} \bigg] \, O_{il} O^{'}_{jk} = - 2i \, \sum_{(ijk)} {\boldsymbol F}_{ijk} \sum_{l} (\delta_{lj}-\delta_{lk}) \, O_{il} O^{'}_{jk} \,,
\end{align}
where $O_{il},O^{'}_{jk}$ are symmetric tensors.
We then obtain
\begin{align}
    {} & \frac{1}{2}\left[{\boldsymbol z}^{(1)},{\boldsymbol z}^{\dagger,\,(1)}\right] = -i \sum_{(ijk)} {\boldsymbol F}_{ijk} \sum_{l} (\delta_{lj}-\delta_{lk}) z^{(1)}_{\{i,l\}} z^{(1),*}_{\{j,k\}} \notag\\
    ={} & -i \sum_{(ijk)} {\boldsymbol F}_{ijk} \sum_{l} (\delta_{lj}-\delta_{lk}) \bigg( \frac1{\eps^4} + \frac{L_{il} + L_{jk} + i \pi (\lambda_{il} - \lambda_{jk} )}{\ep^3} \nonumber \\
    {} & ~~~ + \frac{ L_{il} L_{jk} + \pi^2 \lambda_{il} \lambda_{jk} + i \pi (\lambda_{il} L_{jk} - L_{il} \lambda_{jk}  ) }{\ep^2} \bigg)\notag \\
    ={} & -i \sum_{(ijk)} {\boldsymbol F}_{ijk} \sum_{l} (\delta_{lj}-\delta_{lk}) \bigg( \frac{L_{il} + i \pi \lambda_{il}}{\ep^3}  + \frac{ i \pi (\lambda_{il} L_{jk} - L_{il} \lambda_{jk}  ) }{\ep^2} \bigg)\,,
\end{align}
where we discarded terms that are symmetric in $(il)\leftrightarrow(jk)$ and terms that vanish upon summation because of color conservation.  Finally, thanks to the anti-symmetry of ${\boldsymbol F}_{ijk}$, the $1/\ep^3$ terms sum to zero and we arrive at
\begin{align}
    \label{eqn:renconst-tripole-Z1Z1-lam}
    \frac{1}{2}\left[{\boldsymbol z}^{(1)},{\boldsymbol z}^{\dagger,\,(1)}\right] = {} & \frac{\pi}{\ep^2} \sum_{(ijk)} {\boldsymbol F}_{ijk} \sum_{l} (\delta_{lj}-\delta_{lk}) (\lambda_{il} L_{jk} - L_{il} \lambda_{jk}  ) \nonumber \\
    ={}& \frac{4\pi}{\ep^2} \sum_{(ijk)} {\boldsymbol F}_{ijk} \lambda_{ij} \ln(n_j \cdot n_k) \,.
\end{align}
We also find
\begin{align}
    \label{eqn:renconst-tripole-NLOZ1-lam}
    \frac{1}{2}\left[\tilde{\boldsymbol s}^{(1)}_{\Tau},{\boldsymbol z}^{(1)}-{\boldsymbol z}^{\dagger,\,(1)}\right] ={} &  - \frac{2\pi}{\ep} \sum_{(ijk)} {\boldsymbol F}_{ijk} \sum_{l} (\delta_{lj}-\delta_{lk}) \lambda_{il} \tilde{s}^{(1)}_{\Tau,\{j,k\}} \nonumber \\
    ={} & - \frac{4\pi}{\ep} \sum_{(ijk)} {\boldsymbol F}_{ijk} \lambda_{ij} \tilde{s}^{(1)}_{\Tau,\{j,k\}} \,.
\end{align}

It proves useful to replace $\lambda_{ij}$ in Eqs~(\ref{eqn:renconst-tripole-Z1Z1-lam},~\ref{eqn:renconst-tripole-NLOZ1-lam}) by $\kappa_{ij}$ defined in Eq.~\eqref{eqn:def-kappa}. To do so we note that, for a symmetric tensor $O_{jk}$
\begin{align}
    \sum_{(ijk)} {\boldsymbol F}_{ijk} \lambda_{i,\text{out}} O_{jk}  = 0 \,,
\end{align}
since summation over $(jk)$ involves a product of a fully anti-symmetric and a symmetric tensor.  Similarly,
\begin{align}
    \sum_{(ijk)} {\boldsymbol F}_{ijk} \lambda_{j,\text{out}} O_{jk}  = - \sum_{(jk)} f_{abc} (\colT{j}^a+\colT{k}^a)\colT{j}^b\colT{k}^c \lambda_{j,\text{out}} O_{jk} =0 \,,
\end{align}
where we summed over $i$ using color conservation. Hence, we can simply replace $\lambda\to\kappa$ in the sums in Eqs~(\ref{eqn:renconst-tripole-Z1Z1-lam},\ref{eqn:renconst-tripole-NLOZ1-lam}) and write
\begin{align}
    \label{eqn:renconst-tripole-Z1Z1-kap}
    \frac{1}{2}\left[{\boldsymbol z}^{(1)},{\boldsymbol z}^{\dagger,\,(1)}\right] ={} &  -\frac{4\pi}{\ep^2} \sum_{(ijk)} {\boldsymbol F}_{ijk} \kappa_{jk} \ln(n_i \cdot n_j) \,, \\
    \label{eqn:renconst-tripole-NLOZ1-kap}
    \frac{1}{2}\left[\tilde{\boldsymbol s}^{(1)}_{\Tau},{\boldsymbol z}^{(1)}-{\boldsymbol z}^{\dagger,\,(1)}\right] ={} &  \frac{4\pi}{\ep} \sum_{(ijk)} {\boldsymbol F}_{ijk} \kappa_{jk} \tilde{s}^{(1)}_{\Tau,\{i,j\}} \,.
\end{align}

Proceeding in complete analogy with the procedure described in Section~\ref{sec:NJettiness-tripole}, we can write the Laplace transform of the NLO bare soft function, $\tilde{s}^{(1)}_{\Tau,\,\{i,j\}}$, entering the renormalization counterterm in Eq.~\eqref{eqn:renconst-tripole-NLOZ1-lam} as

\begin{align}\label{eq:tsij_parametrization}
    \tilde{s}^{(1)}_{\Tau,\,\{i,j\}} (u) &= -(4\pi)^{2-\eps} e^{\eps\gamma_E}  \int_{0}^{\infty} \mathrm{d} \tvar \, e^{-u \tvar} \int  [ {\rm d} q] \, S_{ij}(q) \, \delta(\tvar-\Tau(q)) \notag \\
    &= \frac{e^{\eps\gamma_E}}{\Gamma(1-\eps)} \frac{\bar{u}^{2\ep}\, e^{-2\gamma_E\ep}\Gamma(1-2\ep)}{\ep} \left(\frac{n_i\cdot n_j}{2}\right)^{\ep} \int_{0}^{1} {\rm d} \xi \int [{\rm d} \Omega^\perp]_\eps \notag \\ & \phantom{aa} \times \xi^{-1+\ep} \bigg( \left(t_{N,\{ij\}}^{+}(\xi,\phi_{q})\right)^{2\eps} + \left(t_{N,\{ij\}}^{-}(\xi,\phi_{q})\right)^{2\eps} \bigg)\,.
\end{align}

With the above considerations, the renormalized tripole contribution to the soft function in Eq.~\eqref{eq:triple_renorm-0} becomes
\begin{align}
    \tilde{\boldsymbol s}^{\rm tri.,(2),(R)}_{\Tau} (u)
    &= \sum_{(ijk)} {\boldsymbol F}_{ijk}  \kappa_{jk}
    \bigg[
        \frac{N_\eps}{\eps}  I_{\{ij\},k}(u)
        - \frac{4 \pi}{\eps^2}  \ln(n_i\!\cdot\! n_j)
        + \frac{4 \pi}{\eps} \tilde{s}^{(1)}_{\Tau,\,\{i,j\}} (u)
    \bigg] \nn\\
    &\equiv
    \sum_{(ijk)} {\boldsymbol F}_{ijk}  \kappa_{jk}
    \tilde{s}^{\rm tri.,(2),(R)}_{\Tau,\{i,j,k\}}(u) \; .
\end{align}
In general, $\tilde{s}^{\rm tri.,(2),(R)}_{\Tau,\{i,j,k\}}(u)$ admits a Laurent $\ep$-expansion  of the form
\begin{equation}
    \tilde{\boldsymbol s}^{\rm tri.,(2),(R)}_{\Tau,\{i,j,k\}}(u)
    =
    \sum_{n=0}^{3}
    \frac{1}{\eps^n}
    \tilde{s}^{\rm tri.,(2,-n),(R)}_{\Tau,\{i,j,k\}}(u)
    + \mathcal{O}(\eps)\, .
\end{equation}

In the following we discuss the cancellation of the $\ep$-poles. We note  that such cancellations can occur only after summing over all tripole contributions $\{i,j,k\}$. To this end, we recall the useful identities
\begin{equation}\label{eq:sum_properties}
    \sum_{(ijk)} {\boldsymbol F}_{ijk}  \kappa_{jk}  = 0,
    \qquad
    \sum_{(ijk)} {\boldsymbol F}_{ijk}  \kappa_{jk} A_{ijk} = 0\, ,
\end{equation}
for any matrix $A_{ijk}$ symmetric under the exchange of the indices $j$ and $k$, i.e.\ $A_{ijk}=A_{ikj}$. From these relations it immediately follows that the tripole $1/\eps^3$ pole vanishes, since $\tilde{ s}^{\rm tri.,(2,-3),(R)}_{\Tau,\{i,j,k\}}(u)$ is a rational constant and therefore independent of the indices.

After discarding contributions that vanish trivially upon summation, the double-pole contribution becomes
\begin{equation}
 \sum_{(ijk)} {\boldsymbol F}_{ijk}  \kappa_{jk}
 \tilde{s}^{\rm tri.,(2,-2),(R)}_{\Tau,\{i,j,k\}}(u)
 =
 \frac{2}{3}\pi
 \sum_{(ijk)} {\boldsymbol F}_{ijk}  \kappa_{jk}
 \ln \frac{n_i\cdot n_j \, n_i\cdot n_k}{n_j\cdot n_k}
 = 0 \, ,
\end{equation}
where the last step follows from the second identity in Eq.~\eqref{eq:sum_properties}.
Finally, the non-trivial part of the single pole contribution reads
\begin{align}\label{eq:s_ijk_single_pole}
 \tilde{s}^{\rm tri.,(2,-1),(R)}_{\Tau,\{i,j,k\}}&(u)
  \\ 
 & \hspace{-1cm} =
 \int_0^1 {\rm d}\xi \int_0^\pi {\rm d}\phi_{q} \Bigg\{ 2\, \frac{\ln w_k(1,\xi,0,\phi_q) + \ln w_k(\xi,1,0,\phi_q)  + \ln\frac{ (n_i\cdot n_j)^2}{n_i\cdot n_k\,n_j\cdot n_k}}{\xi} \notag \\
 & \hspace{-2.2cm}- \frac{1}{3}\bigg[ -2\ln^2(n_i \cdot n_j) + \ln^2 (n_i \cdot n_k) -2 \ln(n_i \cdot n_k)\ln(n_j \cdot n_k) + 4\ln(n_i \cdot n_j)\ln(n_j \cdot n_k)\bigg]\delta(\xi) \Bigg\}
  \notag  \\
   &\hspace{-1cm} = \frac{1}{3}\pi \left[ 2 ( \ln^2(n_i\cdot n_j) + \ln^2(n_i\cdot n_k)) - 4 \ln(n_i\cdot n_j\,n_i\cdot n_k) \ln(n_j\cdot n_k) + 3 \ln^2(n_j\cdot n_k)\right]\,, \notag
\end{align}
where we removed terms proportional to $\ln(2\sin(\phi_{q}))$ since they vanish upon azimuthal integration.
From Eq.~\eqref{eq:s_ijk_single_pole} it follows that
\begin{equation}
    \sum_{(ijk)} {\boldsymbol F}_{ijk}  \kappa_{jk}
 \tilde{s}^{\rm tri.,(2,-1),(R)}_{\Tau,\{i,j,k\}}(u)
 = 0\,,
\end{equation}
i.e. that also the single pole vanishes.
Finally, we can write the finite remainder of the renormalized tripole correction as
\begin{equation}
   \tilde{s}^{\rm tri.,(2),(R)}_{\Tau,\{i,j,k\}}(u) = \tilde{ s}^{\rm tri.,(2,0),(R)}_{\Tau,\{i,j,k\}}(u) + {\cal O}(\epsilon)\,,
\end{equation}
resulting in the expression in Eq.~\eqref{eq:triple_renorm}.

\section{Laplace Transforms}\label{app:transform}

We define the Laplace transform using the operator
\beq
  \tilde{f}(u)\equiv\LT[f](u)= \int_0^\infty \mathrm{d} \tvar \, e^{-u \tvar} f(\tvar)\,,
\eeq
where $u$ is the Laplace conjugate to $\tvar$.  For convenience we introduce
\begin{align}
  L_u &\equiv \ln(u e^{\gamma_E})\,,
\end{align}
with $\gamma_E$ being the Euler–Mascheroni constant.

The plus distributions $\cL_n(\tvar)$ are defined by their action on a smooth
test function $f(\tvar)$ as
\begin{align}
  \int_0^1 \mathrm{d} \tvar \, \cL_n(\tvar)\, f(\tvar)
  &\equiv \int_0^1 \mathrm{d} \tvar\, \frac{\ln^n(\tvar)}{\tvar}\,[f(\tvar)-f(0)]\,.
\end{align}
They appear in the distribution expansion in $\epsilon$ of $\tvar^{-1 + a\epsilon}$ via \eq{plusDexpansion}. Laplace transforming each side of \eq{plusDexpansion} gives
\begin{align}
  \LT[\tvar^{-1+a\epsilon}](u)
  &= \frac{1}{a\epsilon}
   + \sum_{n=0}^\infty \frac{(a\epsilon)^n}{n!}\,
      \LT[\cL_n](u)\,.
  \label{eq:LT-distribution}
\end{align}
The left-hand side can be calculated directly
\begin{align}
    \label{eqn:LT-tauaep}
  \LT[\tvar^{-1+a\epsilon}](u)
  &\equiv \int_0^\infty \mathrm{d} \tvar \, e^{-u \tvar}\, \tvar^{-1+a\epsilon}
   = u^{-a\epsilon} \Gamma(a\epsilon)\,,
\end{align}
and its small $\epsilon$ expansion gives
\begin{align}
  u^{-a\epsilon}\Gamma(a\epsilon)
  &= \frac{1}{a\epsilon}
   - L_u
   + \frac{a\epsilon}{2}\!\left(L_u^2 + \frac{\pi^2}{6}\right)
   - \frac{(a\epsilon)^2}{6}\!\left(L_u^3 + \frac{\pi^2}{2}L_u
      + 2\zeta_3\right)
   + \cO(\epsilon^3)\,.
  \label{eq:LT-direct}
\end{align}
Now, equating \eq{LT-direct} and \eq{LT-distribution} order by order in $a\epsilon$ provides a straightforward dictionary for the Laplace transforms of the plus distributions. We obtain
\begin{align}\label{eq:laplace-rules}
  \LT[\delta(\tvar)](u) &= 1\,,\nn\\[4pt]
  \LT[\cL_0(\tvar)](u) &= -L_u\,,\nn\\[4pt]
  \LT[\cL_1(\tvar)](u) &= \frac{1}{2}\!\left(L_u^2 + \frac{\pi^2}{6}\right)\,,\nn\\[4pt]
  \LT[\cL_2(\tvar)](u)
    &= -\frac{1}{3}L_u^3
       - \frac{\pi^2}{6} L_u
       - \frac{2}{3}\zeta_3\,,\qquad
\end{align}
and higher–$n$ expressions can be obtained analogously but are not needed for the results presented in this work.
Eq.~(\ref{eq:laplace-rules}) provides the dictionary used in the main text to go back and forth
between momentum space and Laplace space for the soft functions at fixed order in $\alpha_s$ and $\epsilon$.

\addcontentsline{toc}{section}{References}

\bibliographystyle{JHEP}
\bibliography{rref}

\end{document}